\documentclass[aps,pra,twocolumn,superscriptaddress,longbibliography]{revtex4-2}

\usepackage{amsmath}
\usepackage{amssymb}
\usepackage[pdftex]{graphicx}
\usepackage{multirow}
\usepackage{siunitx}
\usepackage[normalem]{ulem}

\begin{document}


\title{Microwave spectroscopy of ultracold-sodium least-bound molecular states}



\author{M. Ballu}
\altaffiliation{These authors contributed equally to this work.}
\affiliation{Universit\'e Sorbonne Paris Nord, Laboratoire de Physique des Lasers,
CNRS UMR 7538, 99 av. J.-B. Cl\'ement, F-93430 Villetaneuse, France}

\author{Z. Yao}
\altaffiliation{These authors contributed equally to this work.}
\affiliation{Universit\'e Sorbonne Paris Nord, Laboratoire de Physique des Lasers,
CNRS UMR 7538, 99 av. J.-B. Cl\'ement, F-93430 Villetaneuse, France}

\author{B. Mirmand}
\affiliation{Universit\'e Sorbonne Paris Nord, Laboratoire de Physique des Lasers,
CNRS UMR 7538, 99 av. J.-B. Cl\'ement, F-93430 Villetaneuse, France}
\author{D. J. Papoular}
\affiliation{Laboratoire de Physique Th\'eorique et Mod\'elisation, CNRS UMR 8089, CY Cergy Paris Universit\'e, 95302 Cergy-Pontoise, France}
\author{H. Perrin}
\affiliation{Universit\'e Sorbonne Paris Nord, Laboratoire de Physique des Lasers,
CNRS UMR 7538, 99 av. J.-B. Cl\'ement, F-93430 Villetaneuse, France}
\author{A. Perrin}
\affiliation{Universit\'e Sorbonne Paris Nord, Laboratoire de Physique des Lasers,
CNRS UMR 7538, 99 av. J.-B. Cl\'ement, F-93430 Villetaneuse, France}


\date{\today}

\begin{abstract}
We have performed microwave spectroscopy of sodium least-bound molecular states, improving the precision of the knowledge of their energies at zero magnetic field by almost three orders of magnitude. Our experimental observations give us access also to states submitted to predissociation, a phenomenon where a bound molecular state can naturally decay into the continuum. Our findings are compared to numerical calculations based on the latest interpolation of sodium interaction potentials and show good agreement, with slight discrepancies in the zero-field energy of the molecular states, suggesting a need for small adjustment of the interaction potentials.
\end{abstract}


\maketitle

\section{Introduction}
\label{sec:intro}

In the field of ultracold atoms, the strength of two-body interactions is well captured by a single parameter, the scattering length. The accurate knowledge of the scattering length is crucial for a proper description of the in- and out-of-equilibrium properties of degenerate quantum gases. In the Born-Oppenheimer approximation, scattering properties of two atoms are determined by the interaction potential they experience at short distances. More accurate measurements of the energy of the molecular bound states associated with this potential thus set stronger constraints on its shape, allowing in turn for the improvement of numerical models describing the interaction between atoms, and a refined determination of the scattering length. In this respect, the least-bound states are of particular importance, since the scattering length is extremely sensitive to their energy. They also play a central role in Feshbach resonances~\cite{Chin2010}, where a pair of free atoms is brought to resonance with a molecular bound state, leading to the divergence of the scattering length.

Alkali-metal atoms have a single valence electron and at short distances their interaction depends on the spin state of the joint electron pair, either singlet or triplet. By contrast, at long distances, the hyperfine splitting interaction, resulting from the coupling between the valence electron spin and the nuclear spin of each atom, is dominant and sets the spin structure of a single atom in its ground state. In the intermediate region, these two energy scales are in competition. In the case of sodium, this has dramatic consequences on the least-bound molecular states, as the hyperfine splitting interaction has similar strength compared with the energy difference between the last bound molecular states of the singlet and triplet potentials. For some particular molecular states, this results in a strong mixing between the singlet and triplet last bound states but also to nearby continuum states. It leads to predissociation, where a bound molecular state is coupled to continuum states, hence strongly limiting its lifetime (\cite{landau3:BH1976}, Sec. 90).

Numerous works have measured or computed the energies of sodium bound molecular states relying on laser-induced fluorescence~\cite{Kusch1978,Barrow1984,Li1985,Babaky1989}, two-photon ionization spectroscopy~\cite{Faerbert1994,Faerbert1996,Faerbert1996b,Faerbert1997} or theoretical analysis~\cite{Friedman-Hill1992,Zemke1994,Gutowski1999,Ho2000}. Raman and two-color photoassociation spectroscopy~\cite{Elbs1999,Samuelis2000,Laue2002,Fatemi2002,DeAraujo2003} refined this knowledge with a typical resolution ranging from 10 to \SI{30}{\mega\hertz}. More recently, the precise characterization of Feshbach resonances~\cite{Inouye1998,stenger:PRL1999,Knoop2011} has constrained even more the shape of singlet and triplet interaction potentials. Taking advantage of the improved knowledge of the energies of the least-bound molecular states, a precise determination of the sodium scattering length has been obtained~\cite{Tiesinga1996,vanAbeelen1999,Crubellier1999,Samuelis2000,Knoop2011}.

In this work, we probe the least-bound molecular states of ultracold sodium atoms with microwave spectroscopy, as also recently demonstrated with rubidium atoms~\cite{Maury2023}, improving the accuracy of previous measurements by nearly three orders of magnitude. This allows us to access the Zeeman structure of individual molecular state and deduce their corresponding Landé $g$ factor. The wide range of microwave field amplitudes accessible with our experimental setup~\cite{Ballu2024} gives us access to the ac Zeeman effect for both atomic and molecular states. Such energy displacement can be seen as the magnetic analog of the ac Stark shift or light shift, usually introduced in the dressed-atom approach~\cite{Agosta1989}. We also determine the energy width of the lowest molecular state undergoing predissociation. Finally, we perform numerical calculations taking advantage of the latest interpolation of sodium singlet and triplet interaction potentials~\cite{Knoop2011}. We correctly reproduce our experimental findings provided small energy offsets, whose value could be used to refine the interaction potentials.

The paper is organized as follows. In Sec.~\ref{sec:theory} we give the theoretical elements needed to express sodium molecular state energies and wavefunctions. In Sec.~\ref{sec:one-photon} we present the experimental apparatus and describe microwave photoassociation spectroscopy of truly bound molecular states and of a state submitted to predissociation. Our results are then compared to numerical calculations. In Sec.~\ref{sec:two-photon}, we use a larger field amplitude to investigate two-photon photoassociation as well as the ac Zeeman effect affecting bound molecular states. In Sec.~\ref{sec:conclusion} we summarize and discuss our results. Additional information concerning the numerical calculations, the compensation of the ac Zeeman shift of the atomic states, and the fit of photoassociation spectra is given in the Appendixes.

\section{Hyperfine structure of sodium molecules}
\label{sec:theory}

In this section we give the theoretical tools to understand the microwave photoassociation spectroscopy of least-bound Na$_2$ molecular states. Since we focus on ultracold atoms, we consider only $s$-wave interactions and we do not take into account any rotational energy. After detailing the possible spin states of a pair of Na atoms, we explain their collisional properties using the center-of-mass frame, with $r$ the relative distance between the two atoms of mass $m$.

\subsection{Singlet and triplet interaction potentials}\label{sec:inter_pot}

Sodium atoms in their ground state are characterized by their electronic spin $\hat{\mathbf{s}}$ with $s=1/2$ and their nuclear spin $\hat{\mathbf{i}}$ with $i=3/2$. Hyperfine interaction $\hbar\omega_{\textrm{hfs}}\hat{\mathbf{i}}\cdot\hat{\mathbf{s}}/2\hbar^2$, with $\hbar$ the reduced Planck constant and $\omega_{\textrm{hfs}}\simeq2\pi\times\SI{1771.6}{\mega\hertz}$, lifts degeneracy between the eight possible spin states which organize into two groups characterized by their total spin $\hat{\mathbf{f}}=\hat{\mathbf{s}}+\hat{\mathbf{i}}$, with $f=1$ or $2$ and split in energy by $\hbar\omega_\textrm{hfs}$.  

The spin of a pair of Na atoms involves $8\times8=64$ different states, but in the case of $s$-wave collisions, only the 36 states symmetric in the exchange of the two atoms are relevant due to the symmetrization rules for indistinguishable bosons. Considering the total hyperfine interaction of both atoms
\begin{align}
    \hat{H}_\textrm{hfs}=\frac{\hbar\omega_{\textrm{hfs}}}{2}\left(\frac{\hat{\mathbf{i}}_1\cdot\hat{\mathbf{s}}_1}{\hbar^2}+\frac{\hat{\mathbf{i}}_2\cdot\hat{\mathbf{s}}_2}{\hbar^2}\right),
\end{align}
the corresponding eigenstates $|\{f_1,f_2\};F,m_F\rangle$ can be labeled by the total spin of the pair $\hat{\mathbf{F}}=\hat{\mathbf{f}}_1+\hat{\mathbf{f}}_2$ where $F=0,1,2,3,4$ and $m_F=-F,\dots,F$. They are split into three different manifolds $\{f_1=1,f_2=1\}$, $\{f_1=1,f_2=2\}$ and $\{f_1=2,f_2=2\}$ separated in energy by $\hbar\omega_{\textrm{hfs}}$. 

At short relative distance $r$, the interaction between two Na atoms depends on their total electronic spin $\hat{\mathbf{S}}=\hat{\mathbf{s}}_1+\hat{\mathbf{s}}_2$. The spin $S$ can take the two values $S=0$ or $1$. For singlet states $S=0$, the atoms interact through the $X^1\Sigma_g^+$ potential $V_\textrm{S}(r)$, while triplet states $S=1$ interact through the $a^3\Sigma_u^+$ potential $V_\textrm{T}(r)$. Previous molecular spectroscopy measurements have allowed us to refine the knowledge of these potentials and in particular the energies of their bound states~\cite{Tiesinga1996,Crubellier1999,Elbs1999,Samuelis2000,Laue2002,Fatemi2002,DeAraujo2003,Knoop2011}. The $X^1\Sigma_g^+$ and $a^3\Sigma_u^+$ potentials include the vibrational levels $\nu_\textrm{S}=0,\dots,65$ and $\nu_\textrm{T}=0,\dots,15$, respectively. Above the last bound state lies a continuum of free states which are \textit{de facto} dissociated. In the following, $\psi^\textrm{S}_{\xi}(r)$ [$\psi^\textrm{T}_{\xi}(r)$] refers to the spatial wave function of an eigenstate of the $X^1\Sigma_g^+$ ($a^3\Sigma_u^+$) potential with energy $E^\textrm{S}_{\xi}$ ($E^\textrm{T}_{\xi}$). The subscript $\xi$ is equal to $\nu_\textrm{S}=0,\dots,65$ ($\xi=\nu_\textrm{T}=0,\dots,15$) for bound states. For continuum states, $\xi=k$, where the momentum $k$ characterizes the asymptotic part of the wave function for $r\rightarrow\infty$, which behaves as $\psi^{\textrm{S},\textrm{T}}_k(r)\propto\sin(k r+\delta^{\textrm{S},\textrm{T}}_k))/r$. The phase shift $\delta^{\textrm{S},\textrm{T}}_k$ depends on the inner part of the spatial wave function and the scattering length $a^{\textrm{S},\textrm{T}}_k$ is set by the limit at vanishing momenta $\tan\delta^{\textrm{S},\textrm{T}}_k/k\rightarrow - a^{\textrm{S},\textrm{T}}_k$.

Singlet and triplet states can be conveniently represented by the spin states $|S,I,F,m_F\rangle$, eigenstates of the operators $\hat{\mathbf{S}}^2$, $\hat{\mathbf{I}}^2=\left(\hat{\mathbf{i}}_1+\hat{\mathbf{i}}_2\right)^2$, $\hat{\mathbf{F}}^2$ and $\hat{F}_z$, projection of $\hat{\mathbf{F}}$ along the quantization axis $z$. It is interesting to note that the states $F=1,3$ and 4 are pure triplet states hence $|S=1,I,F,m_F\rangle=|\{f_1,f_2\};F,m_F\rangle$. In the following, they will be referred to as $|F,m_F\rangle$ since there is no ambiguity. In contrast, the $F=0$ subspace has dimension 2 and each hyperfine splitting eigenstate is a linear combination of the singlet state $|0,0,0,0\rangle$ and the triplet state $|1,1,0,0\rangle$. Similarly, for each $m_F=-2,\dots,2$ the $F=2$ subspace has dimension 3, and each hyperfine splitting eigenstate is a linear combination of the singlet state $|0,2,2,m_F\rangle$ and the two triplet states $|1,1,2,m_F\rangle$ and $|1,3,2,m_F\rangle$.

\subsection{Effect of the hyperfine coupling}

Taking into account the interaction between the two atoms, the Hamiltonian of the system in the center-of-mass frame can be written as
\begin{align}
    \hat{H}_1&=\hat{T}+V_\textrm{S}(\hat{r})\hat{P}_\textrm{S}+V_\textrm{T}(\hat{r})\hat{P}_\textrm{T}+\alpha_\textrm{hfs}(\hat{r})\hat{H}_\textrm{hfs}\label{eq:hamiltonian}
\end{align}
where $\hat{P}_\textrm{S,T}$ are projectors onto singlet and triplet states, $\alpha_\textrm{hfs}(r\rightarrow\infty)= 1$ accounts for electronic distortions of the hyperfine interaction for each atom at short distances~\cite{Knoop2011} and the kinetic energy of the relative motion $\hat{T}$  can be split into a radial and an angular part
\begin{align}
    \hat{T} =-\frac{\hbar^2}{m r^2}  \frac{\partial}{\partial r}\left(r^2\frac{\partial}{\partial r}\right)+\frac{\hat{\mathbf{L}}^2}{m r^2}.
\end{align}
with $\hat{\mathbf{L}}$ the total orbital angular momentum of the atom pair. Here we consider only $s$-wave collisions such that we restrict ourselves to $L=0$ states. 

\begin{figure}[t]
    \centering
    \includegraphics[width=\linewidth]{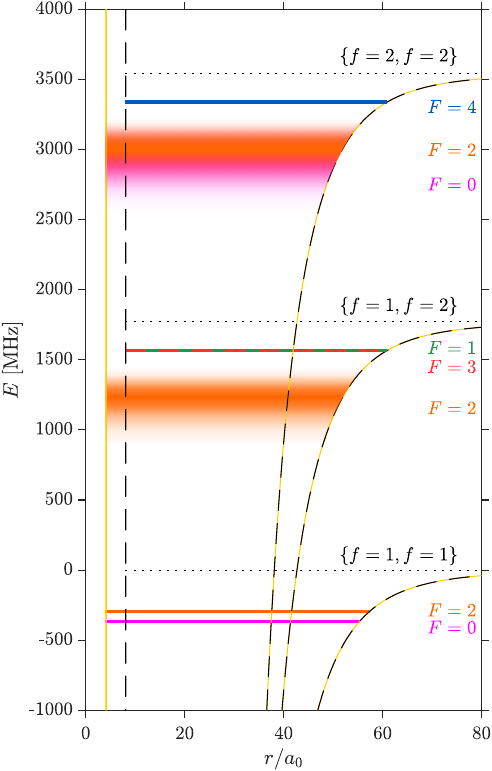}
    \caption{Energy of the Na$_2$ least-bound states (colored lines). Hyperfine interaction distributes these levels between three different manifolds whose dissociation limit is indicated by a black dotted line. The black dashed and yellow solid lines show the triplet interaction potential $V_\textrm{T}(r)$ and singlet interaction potential $V_\textrm{S}(r)$ respectively. The $F=1,3,4$ states are pure triplet states; the $F=0$ and $F=2$ spin states are mixed singlet and triplet states. Within the $\{f=1,f=2\}$ and $\{f=2,f=2\}$ manifolds, this leads to predissociation where the $F=0$ and $F=2$ molecular states can easily leak out to continuum states. The relative distance $r$ is scaled with $a_0$, the Bohr radius.
    } 
    \label{fig:energy_levels}
\end{figure}

The Hamiltonian $\hat{H}_1$ is diagonal in the subspace spanned by the $F=1$, $3$ and $4$ spin states. As mentioned in Sec.~\ref{sec:inter_pot}, these states are pure triplet states, obeying the Hamiltonian $\hat{T}+V_\textrm{T}(\hat{r})+\alpha_\textrm{hfs}(\hat{r})\hat{H}_\textrm{hfs}$. The corresponding eigenstates $|\chi^{F,m_F}_\xi\rangle$ are then easily expressed as 
\begin{align}
    \langle r|\chi^{F,m_F}_\xi\rangle = \phi^{F,m_F}_\xi(r)|F,m_F\rangle,
\end{align}
where $\phi^{F,m_F}_\xi(r)\simeq \psi^\textrm{T}_\xi(r)$, the correction due to $\alpha_\textrm{hfs}$ being very small. Here, $\xi$ is equal to $\nu_\textrm{T}=0,\dots,15$ for bound states and $k$ for continuum states. The last bound state $\nu_\textrm{T}=15$ is represented in Fig.~\ref{fig:energy_levels} for the two manifolds $\{f=1,f=2\}$, with $F=1,3$ and $\{f=2,f=2\}$, with $F=4$.

Within the $F=0$ subspace, singlet and triplet components are coupled through the hyperfine hamiltonian $\hat{H}_\textrm{hfs}$. Below the dissociation limit, this coupling can be treated perturbatively for $\nu_\textrm{S}=0,\dots,64$ and  $\nu_\textrm{T}=0,\dots,14$. Close to dissociation, hyperfine coupling dominates over all other energy scales, so that $\nu_\textrm{S}=65$ and $\nu_\textrm{T}=15$ vibrational states get strongly mixed. Restricting the problem to these two states only, the resulting eigenstates of the Hamiltonian $\hat{H}_1$ have thus a non negligible spin component along both $|\{1,1\};0,0\rangle$ and $|\{2,2\};0,0\rangle$ and their respective spatial wavefunctions are close to a linear combination of $\psi^\textrm{S}_{65}(r)$ and $\psi^\textrm{T}_{15}(r)$. In the following, they will be referred to as $\vert\chi^{\{f,f\};0,0}_{65/15}\rangle$ with $\vert\{f,f\};0,0\rangle$ their main spin component.

The strong mixing imposed by the hyperfine interaction also applies to all continuum states of $V_\textrm{S}$ and $V_\textrm{T}$ in the $F=0$ subspace, labeled by their momentum $k$. While $|\chi^{\{1,1\};0,0}_{65/15}\rangle$ is an isolated state in the energy spectrum (thin magenta line in Fig.~\ref{fig:energy_levels}), $|\chi^{\{2,2\};0,0}_{65/15}\rangle$ lies within mixed continuum states. This gives rise to predissociation where $|\chi^{\{2,2\};0,0}_{65/15}\rangle$ may easily leak out to the continuum. Subsequently its lifetime gets strongly reduced, as illustrated by the wide blurred magenta line in Fig.~\ref{fig:energy_levels}.

In the $F=2$ subspace, a similar treatment can be made for each $m_F=-2,\dots,2$. Below the dissociation limit, the singlet component $|0,2,2,m_F\rangle$ and the two triplet components $|1,1,2,m_F\rangle$ and $|1,3,2,m_F\rangle$ are perturbatively coupled through $\hat{H}_\textrm{hfs}$. Close to dissociation, hyperfine coupling dominates over all other energy scales. Eigenstates of $\hat{H}_1$ are thus superpositions of the three $|\{f_1,f_2\};2,m_F\rangle$ spin states, $\{f_1,f_2\}=\{1,1\}$, $\{1,2\}$ or $\{2,2\}$, with respective spatial wavefunctions which are linear combinations of $\psi^\textrm{S}_{65}(r)$ and $\psi^\textrm{T}_{15}(r)$. In the following, they will be referred to as $|\chi^{\{f_1,f_2\};2,m_F}_{65/15}\rangle$ with $\vert\{f_1,f_2\};2,m_F\rangle$ their main spin component. Similarly to the $F=0$ subspace, while $|\chi^{\{1,1\};2,m_F}_{65/15}\rangle$ is an isolated state in the energy spectrum (thin orange line in Fig.~\ref{fig:energy_levels}), $|\chi^{\{1,2\};2,m_F}_{65/15}\rangle$ and $|\chi^{\{2,2\};2,m_F}_{65/15}\rangle$ lie within mixed continuum states, resulting in predissociation and a short lifetime, as illustrated by the wide blurred orange lines in Fig.~\ref{fig:energy_levels}.

Figure~\ref{fig:energy_levels} summarizes all the results concerning Na$_2$ least-bound states. The origin for the energy scale is set to the $\{1,1\}$ manifold dissociation limit, which corresponds to the energy of two $f=1$ atoms with vanishing relative momentum as in the experiment. As just explained, predissociated states are depicted with a large energy width to account for their limited lifetime.

\subsection{Effect of static and microwave magnetic field}

In the presence of a static magnetic field $\mathbf{B}_s$ or a microwave field $\textbf{B}_\textrm{mw}(t)$ of frequency $\omega$, the Hamiltonian of a pair of Na atoms becomes $\hat{H}(t)=\hat{H}_1+\hat{H}_2(t)$, where
\begin{align}
    \hat{H}_2(t) &= \hat{H}_{\textrm{s}}+\hat{H}_{\textrm{mw}}(t),\\
    \mbox{with }\hat{H}_{\textrm{s}} &= \frac{\mu_B}{\hbar}\left(g_s \hat{\textbf{S}}+g_i \hat{\textbf{I}}\right)\cdot\textbf{B}_\textrm{s}\\
    \mbox{and }\hat{H}_{\textrm{mw}}(t) &= \frac{\mu_B}{\hbar}\left(g_s \hat{\textbf{S}}+g_i \hat{\textbf{I}}\right)\cdot\textbf{B}_\textrm{mw}(t).
\end{align}
Here $\mu_B$ is the Bohr magneton, $g_s\simeq2$ is the Landé $g$ factor of the electronic spin and $g_i\ll g_s$ is the nuclear $g$ factor. Both in the hyperfine basis $|\{f_1,f_2\};F,m_F\rangle$ and in the singlet or triplet basis, the Hamiltonian $\hat{H}_{\textrm{s}}$ is primarily diagonal. Off-diagonal couplings are proportional to $\mu_B g_s B_s$ and can be treated perturbatively as long as they remain small compared with the energy difference between the two coupled states. In this case, $\hat{H}_{\textrm{s}}$ mostly leads to a small shift of each eigenstate energy, yielding a linear Zeeman effect proportional to $\mu_B g_s B_s$.

The effect of the microwave field depends on the frequency $\omega$. When the latter is close to the frequency difference between two eigenstates of $\hat{H}_1$, it leads to coherent Rabi oscillations between them. For off-resonant frequencies and at large microwave amplitude, it also induces significant ac Zeeman shifts on the eigenstates of $\hat{H}_1$.

An accurate numerical treatment of $\hat{H}(t)$ is involved. To estimate the energies of the bound molecular states presented in the subsequent sections, we rely on the following model. We numerically find the eigenstates of $\hat{H}_1$ from the analytical potentials $V_\textrm{S}$ and $V_\textrm{T}$ described in~\cite{Knoop2011} (see Appendix~\ref{sec:app_num_calc} for details). Among the whole set of eigenstates, we keep only the states that are relevant to describe the least-bound molecular states of Na$_2$: the triplet states $|\chi^{F,m_F}_{15}\rangle$ with $F=1$, $3$ and $4$ and $m_F=-F,\dots,F$; the two states $|\chi^{\{1,1\};0,0}_{65/15}\rangle$ and $|\chi^{\{2,2\};0,0}_{65/15}\rangle$ and the states $|\chi^{\{1,1\};2,m_F}_{65/15}\rangle$, $|\chi^{\{1,2\};2,m_F}_{65/15}\rangle$ and $|\chi^{\{2,2\};2,m_F}_{65/15}\rangle$ for $m_F=-2,\dots,2$. We then project $\hat{H}(t)$ on the subspace spanned by these 36 eigenstates. We finally rely on Floquet analysis to compute the least-bound molecular state energy in the presence of a static magnetic field and microwave fields.

In order to characterize the molecular states submitted to predissociation, we rely on a different approach based on coupled-channel calculations, as detailed in Appendix~\ref{sec:quasi_discrete_th}.

\section{Photoassociation spectroscopy of the last Na$_2$ bound states}
\label{sec:one-photon}

We now turn to the experimental outcome of single-photon photoassociation spectroscopy of Na$_2$, starting from a Bose-Einstein condensate (BEC) of $^{23}$Na atoms polarized in the $|f=1,m_f=-1\rangle$ Zeeman state. The spin projection of the atom pair along the quantization axis being $m_F=-2$, the single-photon transitions allowed by selection rules are the molecular states with a spin projection $m_F\in\left\{-3,-2,-1\right\}$ in the states with $F=1$, $2$ or $3$ of the manifolds $\{f=1,f=2\}$ and $\{f=1,f=1\}$. In this section, we present the experimental results of microwave photoassociation spectroscopy for these transitions. 

\begin{figure*}[t]
   \begin{minipage}[c]{.46\linewidth}
      \centering\includegraphics{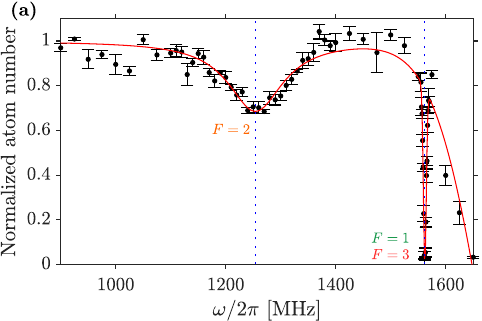}
   \end{minipage} \hfill
   \begin{minipage}[c]{.46\linewidth}
      \centering\includegraphics{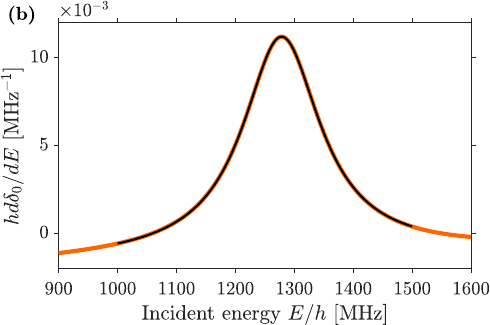}
   \end{minipage}
   \caption{\label{fig:spectro}
(a) Photoassociation spectroscopy of Na$_2$ molecular states of the $\{f=1,f=2\}$ manifold in the presence of a static magnetic field of amplitude $B_s=0.89(1)$\,G. The black circles correspond to the average over three experimental measurements of the remaining atom number after a microwave pulse duration of \SI{80}{\milli\second} and a microwave field amplitude $|B_-|=5.33$\,G for $\omega\simeq 2\pi\times 1254$~MHz. Error bars represent the standard deviation of the different measurements and give an indication of the atom number stability in the experiment. In addition to the atomic hyperfine resonance at \SI{1.77}{\giga\hertz} towards the $f=2$ atomic states, visible for frequencies above \SI{1500}{\mega\hertz}, we observe two molecular resonances: a broad resonance at low frequency and a thin resonance closer to the atomic resonance. The broad resonance, fitted with a Lorentzian function, is centered at \SI{1254.1\pm3.6}{\mega\hertz} and can be attributed to the $F=2$ molecular spin states. Its large width of \SI{172\pm22}{\mega\hertz} at half maximum is due to predissociation (see text). The thin resonance at \SI{1561.9}{\mega\hertz} corresponds to the $F=1,3$ molecular spin states. (b) Numerical characterization of the energy width of $F=2$ molecular states of the $\{f=1,f=2\}$ manifold. The orange line corresponds to the energy derivative of the $s$-wave scattering phase shift $\delta_0$ as a function of the incident energy $E$, assuming a single open channel, namely $|\{1,1\};2,m_F\rangle$. The black line is a Lorentzian function fit supplemented with a linear background representing potential scattering. This leads to a peak energy of $E=h\times \SI{1278}{\mega\hertz}$, slightly detuned compared to the experimental observations and a width of $\gamma=2\pi\times \SI{160}{\mega\hertz}$ in excellent agreement with the experimental data.
}
\end{figure*}

\subsection{Experimental procedure}
\label{sec:procedure}
The microwave spectroscopy of the molecular lines is conducted as follows. We produce a BEC with no visible thermal fraction in a very elongated magnetic trap realized with an atom chip as described in Ref.~\cite{Ballu2024}. The chip design includes a microwave coplanar waveguide (CPW) in the vicinity of which the gas is transported magnetically. At the end of the evaporative cooling procedure, we obtain degenerate gases of typically $10^6$ atoms in the $\vert f=1,m_f=-1\rangle$ Zeeman substate. The confinement is very anisotropic with trapping frequencies  $\omega_{y}\simeq\omega_{z}\simeq 2\pi \times \SI{3.3}{\kilo\hertz}$ and $\omega_{x} \simeq 2\pi \times \SI{6.5}{\hertz}$, where the $x$-axis is the common axis of the CPW and of the main trapping wires on the atom chip. At the bottom of the trap, the atoms experience a local magnetic field $\mathbf{B}_s$ oriented approximately along the $x$-axis, with a typical amplitude $B_s=0.9$\,G that can be increased up to 4.6\,G. The chemical potential of the system $\mu$ is typically lower than $h\times \SI{20}{\kilo\hertz}$ or equivalently $k_B\times\SI{900}{\nano\kelvin}$, and its temperature stays below $\mu/k_B$.

The CPW induces a microwave field $\mathbf{B}_\textrm{mw}(t)=\dfrac{1}{2}\left(\mathbf{\mathcal{B}}e^{-i\omega t}+c.c.\right)$ with $\mathbf{\mathcal{B}}=B_+\mathbf{e}_++B_-\mathbf{e}_-+B_0\mathbf{e}_0$ such that $|B_+|\simeq 1.1|B_-|\gg B_0$, with $\mathbf{e}_+=-(\mathbf{e}_z-i\mathbf{e}_y)/\sqrt{2}$, $\mathbf{e}_-=(\mathbf{e}_z+i\mathbf{e}_y)/\sqrt{2}$, $\mathbf{e}_0=\mathbf{e}_x$. At the position of the atoms, the amplitude $|B_-|$ can be tuned up to 8.35\,G for a microwave frequency around \SI{1.56}{\giga\hertz} (see also Appendix~\ref{sec:app_microwave compensation}). The calibration of these fields is based on the measurement of coherent Rabi oscillations between the Zeeman atomic states $\vert f=1,m_f=-1\rangle$ and $\vert f=2,m_f=-2\rangle$ or $\vert f=1,m_f=0\rangle$ performed at low microwave power and for $\omega\simeq\omega_\textrm{hfs}$, assuming a linear response of the microwave amplifier. Since the transmission of the CPW also depends on $\omega$, we calibrate it relying on a vector network analyzer and take it into account in the estimation of $|B_-|$. The local amplitude of the magnetic field $B_s$ can also be precisely determined from atom loss spectroscopy on the same transitions~\cite{Ballu2024}.

The spectroscopy of Na$_2$ least-bound states is performed by atom loss spectroscopy. The microwave field is switched on at a given power, characterized by an amplitude $|B_-|$ of its $\sigma^-$ component, and for a fixed duration $\tau$. We record the losses induced by the photoassociation of two $\vert f=1,m_f=-1\rangle$ atoms into one of the least-bound Na$_2$ molecular states, while scanning the microwave field frequency $\omega$. The experimental parameters for the different spectra are detailed in Appendix~\ref{sec:app_raw_data}. After the pulse, the atoms are kept in the trap for \SI{440}{\micro\second} before a complete switch off. The molecular states addressed from the initial atomic state are not trapped in the magnetic potential and are then quickly lost in a typical timescale of $\omega_\perp^{-1}\simeq$~\SI{50}{\micro\second}, except for the two states $|\chi^{\{1,1\};2,-2}_{65/15}\rangle$ and $|\chi^{\{1,1\};2,-1}_{65/15}\rangle$ which undergo a magnetic confinement with respective oscillating frequencies $\sqrt{2}$ larger than or identical to the atomic ones. We observe a loss signal for these states as well, which can be attributed to inelastic two-body and three-body collisions among the atoms or between the atoms and the molecules.

\begin{figure*}[t]
   \begin{minipage}[c]{.46\linewidth}
      \centering\includegraphics{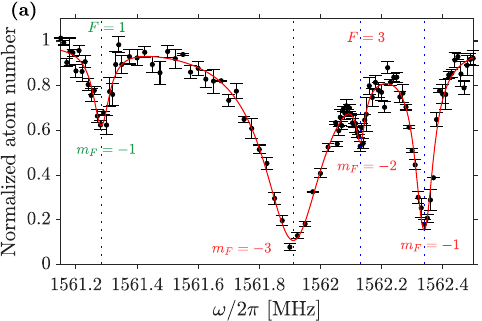}
   \end{minipage} \hfill
   \begin{minipage}[c]{.46\linewidth}
      \centering\includegraphics{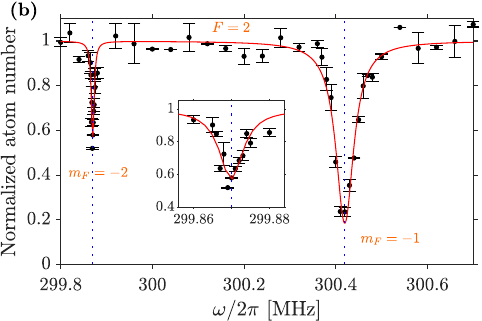}
   \end{minipage}
    \caption{
    Photoassociation spectroscopy of Na$_2$ molecular states in the presence of a static magnetic field of amplitude $B_s=0.89(1)$\,G. The black circles correspond to the average over three experimental measurements of the remaining atom number after a microwave pulse duration of \SI{30}{\milli\second}. Error bars represent the standard deviation of the different measurements and give an indication of the atom number stability in the experiment.
    (a) Photoassociation spectroscopy of the $F=1,3$ molecular spin states of the $\{f=1,f=2\}$ manifold evidencing the Zeeman structure for a microwave field amplitude $|B_-|=0.66$\,G. The four resonances correspond to different $F$, $m_F$ states given in the figure whose coupling is allowed by selection rules. The red line is a fit to a sum of four Lorentzian functions.
    (b) Photoassociation spectroscopy of $F=2$ molecular states of the $\{f=1,f=1\}$ manifold for a microwave field amplitude $|B_-|=1.09$\,G. The very thin resonance at \SI{299.87}{\mega\hertz} is a $\pi$-polarization microwave resonance corresponding to $m_F=-2$, while the broader resonance at \SI{300.42}{\mega\hertz} can be attributed to $m_F=-1$. The red line is a fit to the sum of two Lorentzian functions. The inset shows a close-up of the $m_F=-2$ resonance. Detailed results of the fits for each picture are given in Table~\ref{tab:one_photon_trans} of Appendix~\ref{sec:app_raw_data}.
    } 
    \label{fig:spectrof_1_f_1}
\end{figure*}

\subsection{Observation and characterization of $\{f=1,f=2\}$ predissociated states}
\label{sec:results_prediss}
Fig.~\ref{fig:spectro}(a) shows photoassociation spectroscopy of the $\{f=1,f=2\}$ manifold for $B_s=0.89(1)$\,G. The large peak partially visible on the right part of the plot corresponds to the atomic hyperfine resonance at \SI{1.77}{\giga\hertz} to the $\vert f=2,m_f\rangle$ spin states with $m_f=-2,-1,0$. While these states are untrapped by the magnetic potential, the width of the loss signal is mainly due to the effect of the ac Zeeman shift which expels the atoms from the magnetic trap \cite{Fancher2018}. For $\vert B_-\vert=5.33$\,G and a pulse duration of \SI{80}{\milli\second}, we observe a very broad photoassociation resonance centered at \SI{1254.1\pm 3.6}{\mega\hertz} with a full width at half maximum (FWHM) of \SI{172\pm22}{\mega\hertz}, as well as a thin resonance centered at \SI{1561.9}{\mega\hertz}. The first one can be attributed to the predissociated states $\vert\chi_{65/15}^{\{1,2\};2,m_F}\rangle$ and the second one to the $\vert\chi_{15}^{F,m_F}\rangle$ triplet molecular states with $F=1,3$, which we address in Sec.~\ref{sec:results_one-photon}. 

As discussed previously, the large energy width of the $F=2$ resonance reflects the  finite lifetime  of the  resonant two-atom  state, which decays through predissociation (see \cite{landau3:BH1976}, Sec.~90). We  represent this  process using a  quasidiscrete state (see \cite{landau3:BH1976}, Sec.~134), corresponding to  the complex energy  $E_0-i\hbar\gamma/2$. Its real part $E_0$ is above  the dissociation  threshold, and its imaginary part sets the  lifetime $\gamma^{-1}$. Earlier characterizations of this resonance \cite{Elbs1999,Samuelis2000} involved numerical simulations which closely mimicked the experimental conditions, revealing its impact on the observed resonance position and strength (see \cite{Samuelis2000}, Fig.~5). By contrast, we explore here a different approach and characterize in the absence of static or oscillating magnetic fields this quasidiscrete level, whose energy and width are intrinsic parameters which are independent of the experimental details. Our approach relies on the extraction of the energies and widths of the quasidiscrete states from the energy dependence of the phase shift of scattering wavefunctions (see \cite{landau3:BH1976}, Sec.~134) restricted to the spin state basis of $|\{1,1\};2,m_F\rangle$, $|\{1,2\};2,m_F\rangle$ and $|\{2,2\};2,m_F\rangle$, hence limiting the calculation to three coupled channels. 

Instead of calculating the complex energy of the quasidiscrete state directly, we exploit its impact on scattering states with energies $E$ near the energy of the quasidiscrete state. Among their three spatial components, a single channel is open, namely, $|\{1,1\};2,m_F\rangle$. Hence, they are fully characterized by the $s$-wave phase shift, $\delta_0(E)=\delta_0^{(0)}-\arctan[\hbar\gamma/(2(E-E_0))]$ (see \cite{landau3:BH1976}, Sec.~134). Here, the term $\delta_0^{(0)}$ represents potential scattering, and the arctangent accounts for the resonance near the quasidiscrete state. Its energy derivative exhibits a Lorentzian behavior:
\begin{align}
  \label{eq:delta0deriv}
  \frac{d\delta_0}{dE}
  =P_0^{(0)}+\frac{\hbar\gamma/2}{(E-E_0)^2+\hbar^2\gamma^2/4}
  \ ,
\end{align}
where $P_0^{(0)}(E)=d\delta_0^{(0)}/dE$ weakly depends on $E$.

Further details of the coupled--channel calculation are given in Appendix~\ref{sec:quasi_discrete_th}. Fig.~\ref{fig:spectro}(b) shows the $d\delta_0/dE$ that we extract from the results. We fit to it Eq.~\eqref{eq:delta0deriv}, assuming $P_0^{(0)}(E)$ is linear. We find the width $\gamma=2\pi\times\SI{160}{\mega\hertz}$,
in excellent agreement with the experimental result \SI{172\pm22}{\mega\hertz}.
The predicted resonance energy $E_0$ satisfies
$E_0/h=\SI{1278}{\mega\hertz}$, 
slightly shifted compared to the experimental observation \SI{1254.1\pm3.6}{\mega\hertz}. 

In Appendix~\ref{sec:quasi_discrete_th}, the same treatment is applied to the other two molecular states submitted to predissociation, $|\chi^{\{2,2\};0,0}_{65/15}\rangle$ and $|\chi^{\{2,2\};2,m_F}_{65/15}\rangle$, which we have not investigated experimentally in this work.

\subsection{Observation of individual photoassociation lines}
\label{sec:results_one-photon}

Reducing the microwave field amplitude to $\vert B_-\vert=0.66$\,G with a pulse duration of \SI{30}{\milli\second} allows us to resolve the complete Zeeman structure of the $\vert\chi_{15}^{F=1,3}\rangle$ resonance as shown in Fig.~\ref{fig:spectrof_1_f_1}(a). We observe four resonances corresponding to the molecular states $\vert\chi_{15}^{1,-1}\rangle$, $\vert\chi_{15}^{3,-3}\rangle$, $\vert\chi_{15}^{3,-2}\rangle$ and $\vert\chi_{15}^{3,-1}\rangle$. The coupling to other $\vert\chi_{15}^{F,m_F}\rangle$ molecular states is forbidden by selection rules. Fitting each resonance by a Lorentzian function allows us to determine their center frequency. Note that at resonance $\hbar\omega_0=\Delta E_0$ corresponds to the energy difference between the initial two-atom state and the molecular state, which can both be shifted by the Zeeman effect.

\begin{figure*}[t]
   \begin{minipage}[c]{.46\linewidth}
      \centering\includegraphics{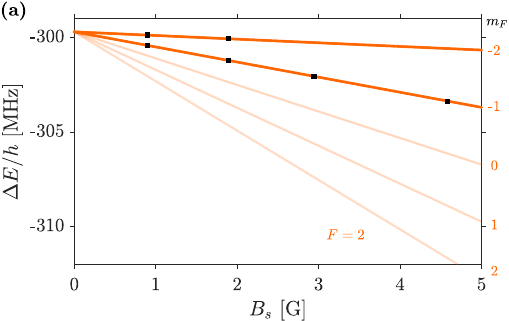}
   \end{minipage} \hfill
   \begin{minipage}[c]{.46\linewidth}
      \centering\includegraphics{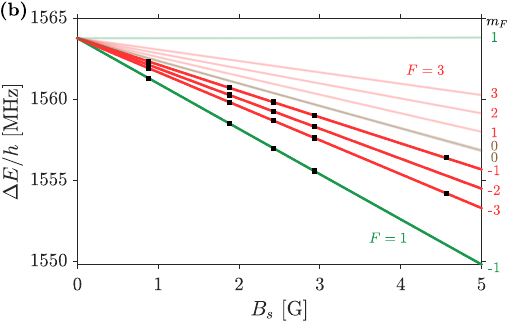}
   \end{minipage}
   \caption{\label{fig:energy_vs_B}
Photoassociation spectroscopy for different values of the static magnetic field $B_s$. Black squares correspond to experimental measurements based on a Lorentzian fit of each resonance (see Table~\ref{manon_data} of Appendix~\ref{sec:app_raw_data} for the microwave pulse parameters used for each point and the detailed fit results). (a) Plot of the $F=2$ molecular states of the $\{f=1,f=1\}$ manifold. The orange lines correspond to numerical calculations shifted by \SI{-6.13}{\mega\hertz}.  (b) Plot of the $F=1,3$ molecular states of the $\{f=1,f=2\}$ manifold. The colored lines correspond to numerical calculations shifted by \SI{-2.97}{\mega\hertz}. The brown color of the $m_F=0$ states indicates the strong mixing of $\vert\chi_{15}^{1,0}\rangle$ and $\vert\chi_{15}^{3,0}\rangle$ by off-diagonal elements of $\hat{H}_s$ (see the text for details).
}
\end{figure*}

The same procedure allows us to observe the Zeeman structure of $F=2$ molecular states of the $\{f=1,f=1\}$ manifold near $\omega\simeq 2\pi\times\SI{300}{\mega\hertz}$ as shown in Fig.~\ref{fig:spectrof_1_f_1}(b). We observe two resonances that we attributes to $\vert\chi_{65/15}^{\{1,1\};2,-2}\rangle$ and $\vert\chi_{65/15}^{\{1,1\};2,-1}\rangle$ which are the only molecular states that can be reached from the initial two-atom state with a single-photon transition because of selection rules. 

The width and depth of each of these resonances mainly reflect the strength of the coupling which is set by the matrix elements of $\hat{H}_\textrm{mw}(t)$ between the two-atom state and the molecular state multiplied by their spatial wave function overlap. As the microwave field amplitude depends on the distance to the CPW~\cite{Ballu2024}, we expect an additional broadening of the order of 10\%. Within the atom trap, the magnetic field amplitude and orientation slightly vary around $B_\textrm{s}$ and $\mathbf{e}_x$. If the energy dependence of the initial atomic state with the static magnetic field is different from the one of the molecular state, this results in an inhomogeneous broadening of the resonance. This effect typically corresponds to a fraction of the chemical potential $\mu$. It also sets a lower limit on the resonance width at low microwave field amplitude. This applies for instance to the $\pi$ lines towards $\vert\chi_{15}^{3,-2}\rangle$ and $\vert\chi_{65/15}^{\{1,1\};2,-2}\rangle$ since $|B_0|\ll|B_{-}|$.

\subsection{Zero field energy and Zeeman shifts}
\label{sec:Zeeman}

In order to access the zero-magnetic-field energy of these molecular states and their Landé $g$ factor, we have repeated the measurement of the photoassociation resonance frequency for different values of the static magnetic field $B_s$ at the bottom of the magnetic trap. Note that tuning this parameter also modifies the trapping frequencies and atom number in the trap. 

The measurements are presented in Fig.~\ref{fig:energy_vs_B} and the parameters deduced from the fits are given in Table~\ref{manon_data} of Appendix~\ref{sec:app_raw_data}. We have compared our results to the model presented in Sec.~\ref{sec:theory} and detailed in Appendix~\ref{sec:app_num_calc}. This model reproduces well the Landé $g$ factor of each individual molecular Zeeman state. We observe, however, that the energy obtained from numerical calculations at $B_s=0$ based on the analytical expression for $V_\textrm{S}$ and $V_\textrm{T}$ described in~\cite{Knoop2011} need to be slightly shifted to reproduce our results. The numerical resolution gives \SI{-293.5\pm 0.5}{\mega\hertz} for the energy of $\vert\chi_{65/15}^{\{1,1\};2,m_F}\rangle$ at $B_s=0$, while our experimental results lead to \SI{-299.66\pm0.02}{\mega\hertz} (see Table~\ref{tab:results}). Moreover, it predicts \SI{1566.8\pm 0.2}{\mega\hertz} for the energy of $\vert\chi_{15}^{F,m_F}\rangle$ with $F=1,3$ at $B_s=0$ while our experimental results lead to \SI{1563.81\pm0.02}{\mega\hertz}. To estimate the accuracy of the numerical calculations, we have investigated the dependence of the molecular state energy with the number of points in the spatial grid (see Appendix~\ref{sec:app_num_calc}).

\begin{figure*}[t]
   \begin{minipage}[c]{.46\linewidth}
      \centering\includegraphics{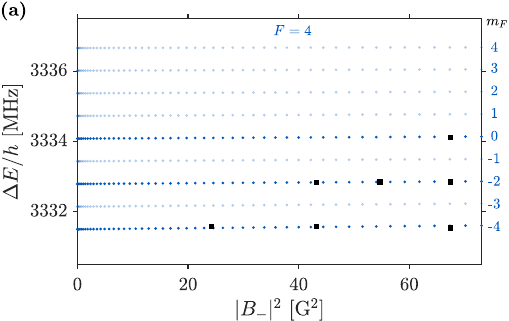}
   \end{minipage} \hfill
   \begin{minipage}[c]{.46\linewidth}
      \centering\includegraphics{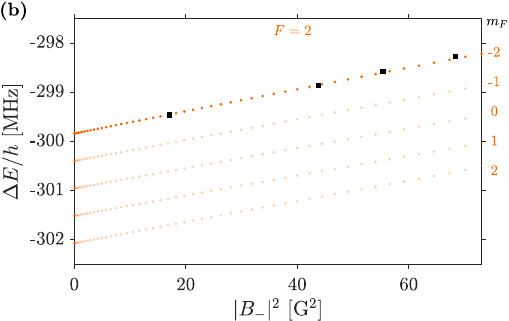}
   \end{minipage}
   \caption{\label{fig:energy_vs_Bmw}
Two-photon photoassociation spectroscopy for different values of the microwave field amplitude $\vert B_-\vert$ and with $B_s=0.92(1)$\,G. Black squares correspond to experimental measurements based on a Lorentzian fit of each resonance (see Table~\ref{tab:2photons} of Appendix~\ref{sec:app_raw_data}). The strong ac Zeeman shift of the atomic transition is compensated by relying on a second microwave field, as explained in Appendix~\ref{sec:app_microwave compensation}. (a) Plot of the  $F=4$ molecular states of the $\{f=2,f=2\}$ manifold. The blue points correspond to numerical calculations. An offset of \SI{-2.99}{\mega\hertz} has been added to the calculated values. (b) Plot of the $F=2$ molecular states of the $\{1,1\}$ manifold. The orange points correspond to numerical calculations. An offset of \SI{-6.13}{\mega\hertz} has been added to the calculated values.
}
\end{figure*}

The uncertainty in the experimental value of the zero-magnetic-field energy of the molecular states is ultimately limited by the inhomogeneous magnetic trapping of the atoms and by collisional shifts between atoms~\cite{Ballu2024} or between atoms and molecules~\cite{Maury2023}. A precise calibration of these effects goes beyond the scope of this paper. We estimate that it is bound by the typical chemical potential of the system $\mu/h\simeq\SI{20}{\kilo\hertz}$. The uncertainty deduced from the fit of the spectroscopy spectra, of the order of a few kilohertz, does not limit the final precision.

We now turn to the estimation of the Land\'e $g$ factor of the molecular states. Since $\vert\chi_{15}^{F,m_F}\rangle$ with $F=1,3$ are pure triplet states, their energy dependence with $B_s$ when $m_F\neq 0$ is directly given by the diagonal elements of $\hat{H}_s$ in the $\vert \{f_1,f_2\};F,m_F\rangle$ basis because a second-order Zeeman shift is negligible in this case. The Zeeman shift can be expressed as $g_{\{f_1,f_2\}}^{(F)} m_F\mu_B B_s$, where
\begin{align}
    g_{\{1,2\}}^{(1)} &= \frac{1}{2}\left(g_i+g_s\right)\simeq 1,\\
    g_{\{1,2\}}^{(3)} &= \frac{1}{12}\left(11 g_i+g_s\right)\simeq \frac{1}{6}.
\end{align}
The states $\vert\chi_{15}^{1,0}\rangle$ and $\vert\chi_{15}^{3,0}\rangle$ are exactly degenerate at $B_s=0$ and their corresponding diagonal elements in $\hat{H}_s$ are also zero. Off-diagonal couplings in $\hat{H}_s$ lift this degeneracy and strongly mix $\vert\chi_{15}^{1,0}\rangle$ and $\vert\chi_{15}^{3,0}\rangle$ (see also~\cite{Maury2023}). 

The energy dependence of $\vert\chi_{65/15}^{\{1,1\};2,m_F}\rangle$ with $B_s$ reflects its spin decomposition in the basis \{$\vert \{1,1\};2,m_F\rangle$, $\vert \{1,2\};2,m_F\rangle$ and $\vert \{2,2\};2,m_F\rangle$\}, since each of these spin components presents a different energy dependence with $B_s$:
\begin{align}
    g_{\{1,1\}}^{(2)} &= \frac{1}{4}\left(5g_i-g_s\right)\simeq -\frac{1}{2},\\
    g_{\{1,2\}}^{(2)} &= \frac{1}{6}\left(5 g_i+g_s\right)\simeq \frac{1}{3},\\
    g_{\{2,2\}}^{(2)} &= \frac{1}{4}\left(3 g_i+g_s\right)\simeq \frac{1}{2}.
\end{align} 
The results of the numerical calculations shown in Fig.~\ref{fig:energy_vs_B}(a) reproduce well our experimental observations. This means that the model correctly captures the weight of the three spin components despite the slight shift at $B_s=0$ mentioned above.

In these studies, we have completely neglected possible ac Zeeman shifts in the energy of these states due to the non zero value of the microwave field amplitude during the spectroscopy. Relying on our numerical model, we have estimated their amplitudes for $\vert B_-\vert=0.6$\,G and $\vert B_+\vert=1.1\vert B_-\vert$: They are of the order of \SI{250}{\hertz} for $\vert\chi_{65/15}^{\{1,1\};2,m_F}\rangle$, \SI{2}{\kilo\hertz} for $\vert\chi_{15}^{1,m_F}\rangle$ and \SI{1}{\kilo\hertz} for $\vert\chi_{15}^{3,m_F}\rangle$ molecular states, one order of magnitude below the experimental uncertainty.

\section{Two-photon photoassociation spectroscopy}
\label{sec:two-photon}

Due to selection rules, in order to perform the microwave spectroscopy of other molecular states, it is necessary to rely on multiple-photon transitions. This requires large microwave field amplitudes and in turn results in significant ac Zeeman shifts on the atomic and molecular states energies. For $\vert B_-\vert$ above a few gauss, we have observed that ac Zeeman shifts due to the atomic transition exceed the chemical potential of the atoms even for a detuning $\delta=\omega-\omega_\textrm{hfs}$ larger than a few hundred of megahertz. The equilibrium position of the atoms in the magnetic potential is then significantly displaced resulting in large excitations of the cloud during the microwave pulse. At the largest microwave field amplitudes, the ac Zeeman shift becomes so strong that the atoms are not trapped anymore. Nevertheless, it is possible to completely compensate this effect at first order relying on a second microwave field of frequency $\omega_c$ with the same amplitude and symmetric with respect to the hyperfine transition, such that $\delta_c=\omega_c-\omega_\textrm{hfs}=-\delta$. Mixing two microwave signals with such characteristics in the CPW, we have experimentally checked the reliability of this technique (see Appendix~\ref{sec:app_microwave compensation} for technical details). This allows us to reach a microwave field amplitude of $\vert B_-\vert\simeq 8.21$\,G for a microwave frequency around \SI{1.66}{\giga\hertz} without visible distortion of the trapping potential.

Two-photon or Raman spectroscopy can then be performed either with two photons having the same frequency $\omega$, or with two photons of frequencies $\omega$ and $\omega_c$. Setting $\delta_c=2\pi\times\SI{100}{\mega\hertz}$ and scanning $\delta\simeq-\delta_c$ over a few megahertz, we first investigated the two-photon spectroscopy of $\vert\chi_{15}^{4,m_F}\rangle$ molecular states of the $\{f=2,f=2\}$ manifold. For these measurements, the two-photon transition is excited with two photons at $\omega$, while the second microwave field at $\omega_c$ is only here to compensate the strong ac Zeeman shift induced on the atomic transition. Nevertheless, both fields may also induce a significant ac Zeeman shift on transitions between two molecular states. In order to extract the transition frequency in the limit of vanishing microwave amplitude, we have repeated the procedure for different microwave field amplitudes and identical static magnetic field $B_s=0.92(1)$\,G and fitted each resonance with a Lorentzian function. The results are presented in Fig.~\ref{fig:energy_vs_Bmw}(a) (see also Table~\ref{tab:2photons} of Appendix~\ref{sec:app_raw_data} for the complete fit results). We observe three distinct resonances corresponding to the molecular states $\vert\chi_{15}^{4,-4}\rangle$, $\vert\chi_{15}^{4,-2}\rangle$ and $\vert\chi_{15}^{4,0}\rangle$. For the resonance toward $\vert\chi_{15}^{4,-4}\rangle$, two $\sigma^-$ photons contribute. For the resonance toward $\vert\chi_{15}^{4,-2}\rangle$, a $\sigma^-$ photon and a $\sigma^+$ photon contribute. For the resonance towards $\vert\chi_{15}^{4,0}\rangle$, two $\sigma^+$ photons contribute. In principle, other states and other polarization combinations are accessible according to selection rules. However, the small relative amplitude of $\vert B_0\vert$ compared to $\vert B_{+,-}\vert$ does not allow us to reach a sufficient coupling to induce two-photon photoassociation.

We have compared these observations to the predictions of the numerical model. As in Sec.~\ref{sec:Zeeman} for the single-photon transitions, the energy of $\vert\chi_{15}^{4}\rangle$ at $B_s=0$ and $\vert B_{+,0,-}\vert=0$ obtained from the calculations, $h\times\SI{3338.4\pm0.2}{\mega\hertz}$, differs slightly from our experimental observations. Introducing a constant offset in the model and adjusting its value by minimizing the difference between the experimental and numerical data leads to a value of $h\times\SI{3335.37\pm0.02}{\mega\hertz}$ for the zero-field energy of the $\vert\chi_{15}^{4}\rangle$ molecular bound state. The plots in Fig.~\ref{fig:energy_vs_Bmw}(a) take this correction into account. From this measurement, we can deduce the energy difference between the $F=4$ molecular states of the $\{2,2\}$ manifold and the $F=1,3$ molecular states of the $\{1,2\}$ manifold. In units of frequency, it is smaller by \SI{67}{\kilo\hertz} than the hyperfine splitting frequency $\omega_\textrm{hfs}/2\pi$. This slight difference corresponds to the effect of the parameter $\alpha_\textrm{hfs}$ introduced in Eq.~\eqref{eq:hamiltonian}.

Interestingly, we do not observe any dependence of the energies of these three states with the microwave field amplitude. As mentioned above, a large ac Zeeman shift may also be expected for the molecular states at such microwave amplitudes. However, since the $F=4$ molecular states are only coupled to the $F=3$ molecular states and their energy difference is close to $\hbar\omega_\textrm{hfs}$, the microwave field at detuning $\delta_c=-\delta$ compensates exactly the ac Zeeman shift induced by the field at detuning $\delta$, confirming in turn the accuracy of the method. This compensation is also well captured by the numerical calculations.

A similar protocol allows us to perform the two-photon photoassociation spectroscopy of the $F=2$ molecular states of the $\{f=1,f=1\}$ manifold, with one photon absorbed from the microwave at $\delta$ followed by a stimulated emission of a second microwave photon at $\delta_c$ [see Fig.~\ref{fig:energy_vs_Bmw}(b)]. We fix $\delta_c=2\pi\times\SI{-150.5}{\mega\hertz}$ and vary $\delta$. We observe only a resonance to the $\vert\chi_{65/15}^{\{1,1\};2,-2}\rangle$ molecular state. This corresponds to the interference of two processes: an absorption of a $\sigma^-$ photon from the microwave field with a detuning $\delta$ associated with a stimulated emission of a $\sigma^-$ photon into the microwave field with a detuning $\delta_c$ or the same process with $\sigma^+$ photons. Two-photon resonances toward $\vert\chi_{65/15}^{\{1,1\};2,-1}\rangle$ are in principle allowed but rely on one $\pi$ polarized photon, for which the microwave amplitude is weak. Also $\vert\chi_{65/15}^{\{1,1\};2,0}\rangle$ is accessible with a two-photon process; however, the matrix element of the microwave coupling is about 14 times smaller in this case as compared with $\vert\chi_{65/15}^{\{1,1\};2,-2}\rangle$ and our experimental signal-to-noise ratio is not sufficient to observe the resonance, even at the largest microwave amplitudes accessible.

\begin{table*}[tbh]
\begin{center}
\begin{tabular}{lcccccc} 
\hline\hline
 vibrational state & $\{f_1,f_2\}$ & $F$ & Expt.~(MHz) & FG~(MHz) & CC~(MHz) & Previous~(MHz) \\ 
\hline\hline
\multirow{2}{8em}{$\nu_S=65;\nu_T=15$} & $\{1,1\}$ & 0 &  & -366.4(9) & & $-393(21)^\textrm{c}$ \\ 
& $\{2,2\}^\textrm{b}$ & 0 &  & 2961.4(6) & 2908 & $\sim3300^\textrm{c}$ \\
\hline
\multirow{3}{8em}{$\nu_S=65;\nu_T=15$}  & $\{1,1\}$ & 2 & -299.66(2) & -293.5(5) & & -293(10)~\cite{Fatemi2002} \\ 
 & $\{1,2\}^\textrm{b}$ & 2 & 1254.1(36) & 1266.8(3)  & 1278 & $1224(24)^\textrm{a}$ \\ 
 & $\{2,2\}^\textrm{b}$ & 2 &  & 3016.6(2) & 3138  & $\sim3300^\textrm{c}$ \\ 
\hline\hline
$\nu_T=15$ & $\{1,2\}$ & 1,3 & 1563.81(2) & 1566.8(2) & & $1568(10)^\textrm{d}$\\
\hline
$\nu_T=15$ & $\{2,2\}$ & 4 & 3335.37(2) & 3338.4(2) & & $3343(10)^\textrm{d}$ \\ 
\hline

\hline\hline
\end{tabular}
\end{center}
\begin{flushleft}
$^\textrm{a}$ From \cite{Elbs1999}.\\
$^\textrm{b}$ State submitted to predissociation.\\
$^\textrm{c}$ From \cite{Samuelis2000}.\\
$^\textrm{d}$ From \cite{Fatemi2002}.
\end{flushleft}
\caption{Summary of the results concerning the zero-magnetic-field energy of the least-bound molecular states of Na$_2$. The energy reference is set to the dissociation limit of the $\{1,1\}$ manifold. For the mixed $\nu_S=65$ and $\nu_T=15$ states, $\{f_1,f_2\}$ and $F$ refer to the main spin component of the molecular state while the superscript $^\star$ denotes a state submitted to predissociation. The Expt. column gives the experimental results obtained from the fit of the microwave spectroscopy spectra. The FG column shows the numerical results obtained with the Fourier grid method. The CC column displays the results of the coupled-channel calculations. Finally, the Previous column give the best previous experimental result we have found in the literature.}
\label{tab:results}
\end{table*}

\begin{figure}[t]
    \centering
    \includegraphics[width=\linewidth]{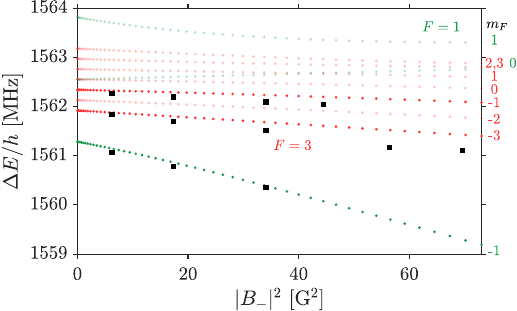}
    \caption{Photoassociation spectroscopy of the $F=1,3$ molecular states of the $\{f=1,f=2\}$ manifold for different values of the microwave field amplitude $\vert B_-\vert$. The static magnetic field is equal to $B_s=0.90(1)$\,G. Black squares correspond to experimental measurements based on a Lorentzian fit of each resonance (see Table~\ref{tab:dressing} of Appendix~\ref{sec:app_raw_data}). The colored points correspond to numerical calculations. The strong ac Zeeman shift of the atomic transition is compensated by relying on a second microwave field as explained in Appendix~\ref{sec:app_microwave compensation}. 
    } 
    \label{fig:f_1_f_2_F_1_F_3_vs_Bmw_2_photons}
\end{figure}

We have also investigated the energy dependence of the single-photon $F=1,3$ molecular resonances with the microwave field amplitude, as shown in Fig.~\ref{fig:f_1_f_2_F_1_F_3_vs_Bmw_2_photons}. In this case, the numerical model reproduces less accurately the experimental data, in particular for $\vert\chi_{15}^{3,-3}\rangle$ and $\vert\chi_{15}^{3,-1}\rangle$. We have checked that the numerical results are sensitive to the characteristics of the $F=2$ states submitted to predissociation, e.g., zero-field energy and overlap of the spatial wavefunctions. These features are probably oversimplified in our model. These limitations call for future improvement in our numerical model.

Finally, we have also tried to look for the $\vert\chi_{65/15}^{\{1,1\};0,0}\rangle$ molecular state, which is accessible through the absorption of a $\sigma^-$ microwave photon and the subsequent emission of a $\sigma^+$ microwave photon. However, the matrix element of the microwave coupling to this state is about 10 times smaller as compared with the coupling to the $\vert\chi_{65/15}^{\{1,1\};2,-2}\rangle$ state and despite our efforts our signal to noise ratio did not allow us to locate the resonance.

\section{Conclusion}
\label{sec:conclusion}

In this paper we have carried out the microwave spectroscopy of Na$_2$ least-bound states and pushed the precision of the determination of the energies of these states by nearly three orders of magnitude compared with previous works. The residual uncertainty comes from the inhomogeneity of the magnetic field in the atom trap and from collisional shifts between atoms but also between atom and molecules. This is responsible for a systematic uncertainty in the determination of the energies of Na$_2$ least-bound states. We estimated that this uncertainty is bounded by the chemical potential of the gas $\mu\simeq h\times\SI{20}{\kilo\hertz}.$~\footnote{This uncertainty is far too large to account for second-order spin-orbit coupling which ultimately limits the lifetime of pure triplet states~\cite{Kokkelmans2001,Vanhaecke2002}, expected to be larger than 100~s.} We have compared our experimental results to numerical calculations, which show good agreement with the experimental data despite small shifts in the zero-magnetic-field energy of the probed states. Due to the relatively large amplitude of the microwave field available on the experiment setup, we were able to measure the energy width of a molecular state submitted to predissociation. These experimental results are in good agreement with coupled-channel calculations that have allowed us to characterize other molecular states submitted to predissociation. At large amplitude of the microwave fields, it is also possible to access specific molecular states with a two-photon transition. The microwave dressing of the molecular states themselves is responsible for an ac Zeeman shift that we have characterized. The main results of the paper are summarized in Table~\ref{tab:results} (see Appendix~\ref{sec:app_low_vib} for a similar Table for $\nu_S=64$ and $\nu_T=14$ vibrational states).

For alkali-metal atoms, the microwave coupling of a scattering state to a molecular bound state was shown to give rise to a Feshbach resonance~\cite{Papoular2010,papoular:PhD2011}, whose frequency width $\Delta\omega$ scales as the square of the microwave field amplitude. For sodium atoms, numerical calculations predict a scaling of $2\pi\times1.4$\,kHz/G$^2$. For the largest microwave amplitude accessible in the experiment, the estimated width should be significantly larger than the energy spread of the gas in the magnetic trap and lead to observable effects on the equilibrium properties of the system due to the modification of the scattering length. These considerations should stimulate dedicated investigations in the vicinity of the molecular resonances.

Strong microwave coupling leads to the mixing of the hyperfine states of the atom, but also of the hyperfine molecular states. As inelastic collisions in a degenerate gas of alkali-metal atoms substantially depend on the spin state of the atoms~\cite{Stamper-Kurn1998,Tojo2009}, a complete characterization of the two- and three-body loss rates of the system in the presence of a large amplitude microwave field and possibly near a microwave-induced Feshbach resonance would constitute an interesting achievement. It relates to a very recent work on the control of the imaginary part of the lithium scattering length with a radiofrequency modulation of a magnetic field near a Feshbach resonance \cite{Guthmann2025}.

\begin{acknowledgments}
We thank E. Luc, A. Orban, N. Bouloufa and O. Dulieu for enlightening discussions on the physics of sodium molecular bound states at an early stage of the project, and for bringing Refs.~\cite{DeAraujo2003} and \cite{Knoop2011} to our attention. We also thank J. Beugnon for a careful reading of the manuscript and B. Laburthe-Tolra for useful comments and bringing Ref.~\cite{Kokkelmans2001} and \cite{Vanhaecke2002} to our attention.   This work was supported by the Agence Nationale de la Recherche under Projects No. ANR-21-CE47-0009-03 and No. ANR-22-CE91-0005-01.
\end{acknowledgments}

\appendix

\section{Numerical calculations}
\label{sec:app_num_calc}

In order to find the eigenstates and eigenenergies of $\hat{H}_1$ defined by Eq.~\eqref{eq:hamiltonian}, we rely on the Fourier grid method~\cite{Dulieu1995}. We use a spatial grid of $10^4$ evenly spaced points between $3 \, a_0$ and $350 \, a_0$ where $a_0$ is the Bohr radius. Since $F=1,3,4$ spin states are eigenstates of $\hat{H}_\textrm{hfs}$ and also pure triplet states, we can restrict the problem to a given $|F,m_F\rangle$ state and write
\begin{align}
    \langle F,m_F|\hat{H}_1|F,m_F\rangle = \hat{T}+V_\textrm{T}(r)+l_F\frac{\alpha_\textrm{hfs}(r)}{4}\hbar\omega_\textrm{hfs},
\end{align}
where $l_1=l_3=-1$ and $l_4=3$. Solving the corresponding Schrödinger equations, we determine the spatial wavefunctions $\langle r|\chi^{F,m_F}_\xi\rangle$ and their energies. Note that these solutions do not depend on the value of $m_F$.

For $F=0$ or $2$, $m_F$ states, we solve instead the coupled-channel system in the subspace spanned by $|\{1,1\};0,0\rangle$ and $|\{2,2\};0,0\rangle$ or by $|\{1,1\};2,m_F\rangle$, $|\{1,2\};2,m_F\rangle$ and $|\{2,2\};2,m_F\rangle$. Again, the results do not depend on the value of $m_F$. In all these calculations we rely for $V_\textrm{S}(r)$, $V_\textrm{T}(r)$ and $\alpha_\textrm{hfs}(r)$ on the analytical expressions given in~\cite{Knoop2011}. We observe that the results depend on the number of points of the spatial grid. We estimate the uncertainty of the results by repeating the calculation for different grid sizes up to $10^4$ points. For a given molecular bound-state energy, we fit the set of results at different grid sizes by a decreasing exponential plus an offset. We define the uncertainty on the energy of this molecular bound state as the difference between the fitted offset and the result for a grid of $10^4$ points.

\begin{figure*}[t]
   \begin{minipage}[c]{.46\linewidth}
      \centering\includegraphics{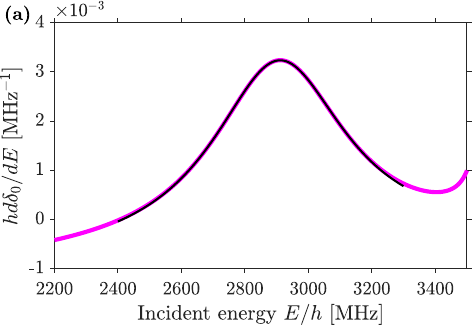}
   \end{minipage} \hfill
   \begin{minipage}[c]{.46\linewidth}
      \centering\includegraphics{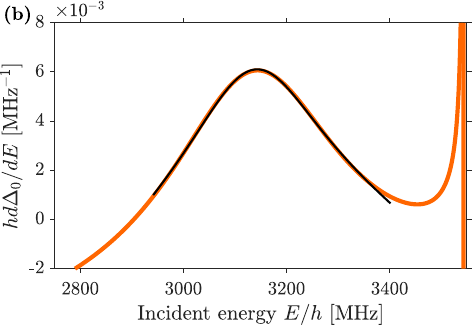}
   \end{minipage}
   \caption{\label{fig:quasi_dis_f2f2}
(a) Numerical characterization of the energy width of the $F=0$ molecular state of the $\{f=2,f=2\}$ manifold. The purple line corresponds to the energy derivative of the $s$-wave scattering phase shift $\delta_0$ as a function of the incident energy $E$, assuming a single open channel, namely $|\{1,1\};0,0\rangle$. The black line is a Lorentzian function fit supplemented with a linear background representing potential scattering. This leads to a peak energy of $E=h\times \SI{2908}{\mega\hertz}$ and a width of $\gamma=2\pi\times \SI{516}{\mega\hertz}$. (b) Numerical characterization of the energy width of $F=2$ molecular states of the $\{f=2,f=2\}$ manifold. From the $S$ matrix defined on the subspace spanned by the two open channels $|\{1,1\};2,m_F\rangle$ and $|\{1,2\};2,m_F\rangle$, we define the phase $\Delta_0$ as $\det(S)=\exp(2i\Delta_0)$. The orange line in the plot corresponds to the energy derivative of $\Delta_0$ as a function of the incident energy $E$. The black line is a Lorentzian function fit supplemented with a quadratic background. This leads to a peak energy of $E=h\times \SI{3138}{\mega\hertz}$ and a width of $\gamma=2\pi\times \SI{324}{\mega\hertz}$.
}
\end{figure*}

Among all the numerical solutions $|\chi^{\{f_1,f_2\};F,m_F}_\xi\rangle$, we identify the ones corresponding to the least-bound states of Na$_2$ molecules. For $F=1,3,$ and $4$, it simply corresponds to $|\chi_{15}^{F,m_F}\rangle$. For $F=0$ and $2$, it is relatively easy to identify $|\chi_{65/15}^{\{1,1\};0,0}\rangle$ and $|\chi_{65/15}^{\{1,1\};2,m_F}\rangle$ since they are pure bound states and isolated in energy compared with other eigenstates. Since $|\chi_{65/15}^{\{2,2\};0,0}\rangle$, $|\chi_{65/15}^{\{1,2\};2,m_F}\rangle$ and $|\chi_{65/15}^{\{2,2\};2,m_F}\rangle$ are degenerate in energy with continuum states, the numerical solutions correspond to mixtures of each of these states with continuum states or predissociated states. Among those, we pick the ones that lead to the largest overlap with previously identified states. For the results presented in the main text, we have checked that this choice is not very sensitive.

We then project $\hat{H}(t)=\hat{H}_1+\hat{H}_2(t)$ to the subspace spanned by the collection of eigenstates identified as least-bound states of Na$_2$ molecules as we just explained. In this subspace, $\hat{H}_1$ is obviously diagonal. The static part of $\hat{H}_2(t)$, which corresponds to $\hat{H}_s$, has diagonal elements corresponding to the first-order Zeeman energy shift, while off-diagonal coupling is responsible for second-order Zeeman energy shifts. The time-dependent part of $\hat{H}_2(t)$ corresponds to the coupling to the microwave field $\mathbf{B}_\textrm{mw}(t)$. Relying on the Floquet formalism~\cite{Shirley1965}, we transform the Hamiltonian $\hat{H}(t)$ into a time-independent Hamiltonian that can be expressed as the sum of two terms,
\begin{align}
    \hat{H}_\textrm{0} &=\sum_{n=-\infty}^\infty \left(\hat{H}_1+\hat{H}_\textrm{s}+n\hbar\omega\right)|n\rangle\langle n|,\\
    \hat{H}_\textrm{c} &=\sum_{n=-\infty}^\infty \hat{H}_\textrm{mw}|n\rangle\langle n+1|+\hat{H}_\textrm{mw}^\dagger|n+1\rangle\langle n|,
\end{align}
with 
\begin{align}
    \hat{H}_\textrm{mw}(t)=\hat{H}_\textrm{mw}e^{-i\omega t}+\hat{H}_\textrm{mw}^\dagger e^{i\omega t}.
\end{align}
Restricting $\hat{H}_\textrm{0}+\hat{H}_\textrm{c}$ to the Floquet manifolds $n=-3,\dots,3$ and diagonalizing it, we deduce the energies of the molecular states in the presence of the microwave field.

\section{Numerical characterization of resonances at
quasidiscrete levels}
\label{sec:quasi_discrete_th}

In order to characterize numerically the energy width of molecular states submitted to predissociation, we calculate the corresponding scattering wavefunctions using the coupled--channel approach \cite{verhaar:PRA2009}, our c++ implementation of which is described in Ref.~\cite{papoular:PhD2011}, Chap.~12. We use the singlet and triplet potentials $V_{\textrm{S},\textrm{T}}(r)$ for sodium given in Ref.~\cite{Knoop2011} and apply the adiabatic accumulated--phase boundary condition at the radius $r_0=16\,a_0$. We choose the phases at $r_0$ to reproduce the singlet and triplet scattering lengths and calculate their derivatives from the potentials $V_{\textrm{S},\textrm{T}}(r)$ for $r<r_0$. For testing purposes, we supplement the Hamiltonian $\hat{H}_1$ with the Zeeman term $\hat{H}_s$ and check that we reproduce the positions of the four $s$-wave Feshbach resonances known to affect ${}^{23}\mathrm{Na}$ \cite{Knoop2011,stenger:PRL1999} with a relative accuracy less than $1\%$. Subsequently, we perform all calculations in the absence of a magnetic field.

We now describe the results obtained for the two resonances corresponding to the $F=0$ and $F=2$ quasidiscrete states whose energies lie below the dissociation threshold of the $\{f=2,f=2\}$ manifold. They are both mentioned, e.g., in Ref.~\cite{Samuelis2000}, Fig.~5b, but to our knowledge their energies and widths have not yet been accurately determined. As for $|\chi^{\{1,2\};2,m_F}_{65/15}\rangle$ discussed in Sec.~\ref{sec:results_prediss}, we follow the quasidiscrete level formalism~(see \cite{landau3:BH1976}, Sec.~134).

As mentioned in Sec.~\ref{sec:inter_pot}, the $F=m_F=0$ subspace is spanned by the two states $|\{1,1\};0,0\rangle$ and $|\{2,2\};0,0\rangle$. The real part $E_0$ of the eigenvalue of $H_1$ corresponding to the quasidiscrete level lies between the dissociation limit of the $\{1,1\}$ and $\{2,2\}$ manifolds. Hence, the relevant collisions involve a single open channel $|\{1,1\};0,0\rangle$. The same theory as the one discussed in the main text is then directly applicable and the results are illustrated in Fig.~\ref{fig:quasi_dis_f2f2}(a). We find $E_0/h=\SI{2908}{\mega\hertz}$ and $\gamma=2\pi\times\SI{516}{\mega\hertz}$.

For each $m_F$, the $F=2$ subspace has dimension 3 and can be spanned by $|\{1,1\};2,m_F\rangle$, $|\{1,2\};2,m_F\rangle$ and $|\{2,2\};2,m_F\rangle$. The real part $E_0$ of the eigenvalue of $H$
corresponding to the quasidiscrete level lies now between the dissociation limit of the $\{1,2\}$ and $\{2,2\}$ manifolds. Hence, the relevant collisions involve two open channels $|\{1,1\};2,m_F\rangle$ and $|\{1,2\};2,m_F\rangle$. They are characterized by a $2\times 2$ unitary scattering matrix $S$, whose determinant $\det(S)=\exp(2i\Delta_0)$ has modulus 1. The resonant behavior is described by
Eq.~(\ref{eq:delta0deriv}) where $\delta_0$ is replaced by $\Delta_0$ (see \cite{landau3:BH1976}, Sec.~145). We fit it to the $d\Delta_0/dE$ obtained from our numerical coupled--channel calculations, with  $P_0^{(0)}(E)$ quadratic. We find $E_0/h=\SI{3138}{\mega\hertz}$ and $\gamma=2\pi\times\SI{324}{\mega\hertz}$. 

\section{ac Zeeman shift compensation}
\label{sec:app_microwave compensation}

In a two-level approximation, the ac Zeeman shift induced in the $|f=1,m_f=-1\rangle$ to $|f=2,m_f=-2\rangle$ atomic transition by the microwave field can be approximated by $6|\Omega_-|^2/4\delta$ where $\Omega_-=-g_s\mu_B B_-/4\hbar$. For the highest microwave amplitude accessible in the experiment, this corresponds to an ac Zeeman shift of about $h\times$\SI{220}{\kilo\hertz}. Since the CPW produces an inhomogeneous microwave field whose amplitude decreases with the vertical distance to the waveguide (see~\cite{Ballu2024}), this results in a gradient which perturbs the magnetic confinement of the atoms. As a result, abruptly switching on the microwave field leads to the transverse excitation of the BEC and hence to atomic losses. For the highest microwave amplitude that we can reach experimentally, all the atoms are lost.

In order to compensate this ac Zeeman shift, we rely on a second microwave field of frequency $\omega_\textrm{c}$ and of equal amplitude $B_{+,-,0}^\textrm{c}=B_{+,-,0}$ but with opposite-sign detuning  $\delta_\textrm{c}=-\delta$. At first order, this completely suppresses any ac Zeeman shift. 

To fine-tune the parameters of this second microwave field for ac-Zeeman-shift compensation, we shine the two strong-amplitude microwave fields onto the atoms for a duration of \SI{1}{\milli\second} and with $\delta_\textrm{c}=-\delta=\SI{206.6}{\mega\hertz}$. After the pulse, we keep the atoms in the trap for \SI{500}{\micro\second} before complete switch off. We then image the atoms after a 10-ms time of flight. We repeat this protocol for different amplitude $B_-^\textrm{c}$, keeping $|B_-|=5.53(8)$\,G constant.

\begin{figure}[t]
    \centering
    \includegraphics[width=\linewidth]{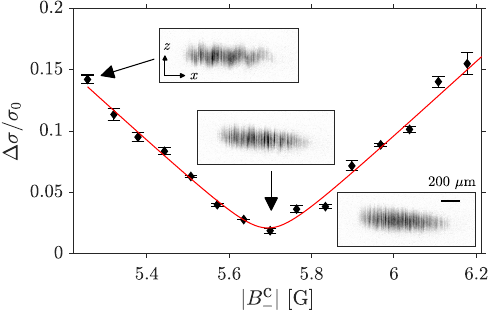}
    \caption{Ratio of the standard deviation $\Delta\sigma$ with the mean transverse width of the cloud $\sigma_0$ for different values of the amplitude of $|B_-^\textrm{c}|$ and for fixed $|B_-|=5.53(8)$\,G. This quantity characterizes the amplitude of the cloud excitation when the ac Zeeman shift is not perfectly compensated (see the text for details). The error bars represent the standard deviation of two measurements. The red line is a guide to the eye. The top left inset shows an absorption imaging picture of the cloud after \SI{10}{\milli\second} time of flight. The snake shape of the cloud is due to the excitation of the transverse dipole mode of the BEC and to the fact that the transverse oscillating frequency varies by about 10\% in the $x$ direction. The middle inset shows an absorption imaging picture of the cloud when the ac Zeeman effect is optimally compensated at $|B_-^\textrm{c}|=5.70$\,G. We attribute the slight difference between $|B_-^\textrm{c}|$ and $|B_-|$ to systematic errors in the microwave field calibration (see the text for details). The bottom right inset shows an absorption imaging picture of the cloud when $|B_-^\textrm{c}|=|B_-|=0$\,G.
    } 
    \label{fig:compensation}
\end{figure}

When the ac Zeeman shift is not perfectly compensated, the transverse dipole mode of the BEC is excited which is reflected in the shape of the cloud after a 10-ms time of flight as shown in the top left inset of Fig.~\ref{fig:compensation}. The snake shape of the cloud in this case comes from the fact that the transverse oscillating frequency is not homogeneous along the longitudinal direction of the gas but varies by about 10\%, which leads to a dephasing in the transverse oscillation. We take advantage of this dephasing to identify the optimal amplitude for the second microwave field to compensate the ac Zeeman shift of the first one. The optimal amplitude is the one that minimizes the dephasing in the transverse oscillation, which we observe in the vertical direction.

In order to quantify the amplitude of the excitation, we first calculate the vertical position of the center of mass of a narrow longitudinal section of the gas at a given position $x$ along the longitudinal axis
\begin{align}
    \overline{z}(x) = \frac{1}{\bar{n}(x)}\int dz \ z \ n(x,z)
\end{align}
where $n(x,z)$ is the two-dimensional density obtained by absorption imaging along the $y$ axis and $\bar{n}(x)=\int dz \ n(x,z)$ is the density integrated along the $z$ axis at position $x$. We then compare $\overline{z}(x)$ to $\overline{z}_0(x)$ obtained for $B_-^\textrm{c}=B_-=0$ by calculating the standard deviation $\Delta\sigma$ of $\bar{z}(x)$ given by
\begin{align}
    \Delta\sigma = \sqrt{\frac{1}{L_x}\int_0^{L_x} dx\left[\overline{z}(x)-\overline{z}_0(x)\right]^2}
\end{align}
where $L_x$ is the longitudinal length of the BEC. 

In the absence of excitation, the rms transverse width of the system $\sigma_0$ for $|B_-^\textrm{c}|=|B_-|=0$\,G is given by
\begin{align}
    \sigma_0 = \sqrt{\frac{1}{N}\int dx dz \ z^2 \ n(x,z)-\left[\frac{1}{N}\int dx dz \ z \ n(x,z)\right]^2}
\end{align}
with $N$ the total atom number in the gas. We show in Fig.~\ref{fig:compensation} the ratio of $\Delta\sigma/\sigma_0$ for different microwave amplitudes of the compensation field $B_-^\textrm{c}$. We observe a clear minimum for $|B_-^\textrm{c}|=5.70$\,G.

We attribute the slight difference between the optimum $|B_-^\textrm{c}|$ and the amplitude $|B_-|$ to systematic errors in the microwave field amplitude calibration. They are calibrated independently by measuring the frequency of coherent Rabi oscillations at resonance with the $|f=1,m_f=-1\rangle$ to $|f=2,m_f=-2\rangle$ atomic transition at low microwave power as described in~\cite{Ballu2024}. We then calibrate each microwave field amplitude at the entrance of the CPW with a spectrum analyzer, for the settings used in Fig.~\ref{fig:compensation} as well as for the one used for the Rabi frequency measurements. Assuming a linear response of the CPW and the complete independence of both fields, we deduce $B_-$ and $B_-^\textrm{c}$ for Fig.~\ref{fig:compensation}.

\begin{table}[b]
\begin{center}
\begin{tabular}{lcccll} 
\hline\hline
 Vibrational state & $S$ & $I$ & $F$ & FG~(MHz)  & Previous~(MHz) \\ 
\hline\hline
\multirow{2}{8em}{$\nu_S=64;\nu_T=14$}  & 0 & 0 & 0 & -11090(13) & -11224(27)$^\textrm{a}$ \\ 
& 1 & 1 & 0  & -5540(4) & \\
\hline

\multirow{3}{8em}{$\nu_S=64;\nu_T=14$} & 0 & 2 & 2 & -10938(9) & -11080(27)$^\textrm{a}$\\
& 1 & 3 & 2  & -6627(8) & -6659(15)$^\textrm{b}$ \\ 
& 1 & 1 & 2 & -4449(8) &  \\
\hline\hline
$\nu_T=14$ & 1 & 1,3 & 1,3 & -5413(2) & -5457(15)$^\textrm{b}$ \\ \hline
$\nu_T=14$ & 1 & 3 & 4 & -3642(2)  & -3669(15)$^\textrm{b}$ \\ 
\hline

\hline\hline
\end{tabular}
\end{center}
\begin{flushleft}
$^\textrm{a}$ From \cite{Elbs1999}.\\
$^\textrm{b}$ From \cite{DeAraujo2003}.
\end{flushleft}
\caption{Summary of the results concerning the energy of the molecular states of Na$_2$ corresponding to $\nu_S=64$ and $\nu_T=14$ vibrational states. The energy reference is set to the dissociation limit of the $\{1,1\}$ manifold as in Table~\ref{tab:results}. For the $\nu_S=64$ and $\nu_T=14$ states perturbatively mixed by the hyperfine interaction, $S$, $I$ and $F$ indicate the main spin component of the molecular state. The FG column shows the numerical results obtained with the Fourier grid method. Finally, the Previous column give the best experimental result we have found in the literature.
\label{tab:low_vib}
}
\end{table}

\section{Energy of the molecular bound states involving the $\nu_S=64$ and $\nu_T=14$ vibrational states}
\label{sec:app_low_vib}

In this appendix we show in Table~\ref{tab:low_vib} the energies of lower bound molecular states corresponding to $\nu_T=14$ or to the perturbative mixing of $\nu_S=64$ and $\nu_T=14$ due to hyperfine interaction.

\section{Experimental parameters and fit results}
\label{sec:app_raw_data}

Tables~\ref{tab:one_photon_trans}-\ref{tab:dressing} present all the relevant experimental parameters and fit results for the microwave photoassociation spectra discussed in the main text. The value of the static magnetic field $B_s$ is calibrated from microwave spectroscopy of the atomic transition~\cite{Ballu2024}. The calibration of the microwave field amplitude is made with the method explained in Appendix~\ref{sec:app_microwave compensation}. All spectra are fitted with a Lorentzian function. All the uncertainties indicated in the tables are deduced from the fit covariance matrix. They do not take into account the systematic uncertainties discussed in the main text.

\begin{table*}[htb]
\centering
\begin{tabular}{lllllll} 
\hline\hline
 \multicolumn{1}{c}{Final main spin state} & Polarization & $B_s$~(G)& $|B_-|$~(G) & $\tau$~[ms] & \multicolumn{1}{c}{$\omega_0/2\pi$~(MHz)} & \multicolumn{1}{c}{FWHM~(MHz)}  \\ 
\hline

\multicolumn{1}{c}{$\vert \{f=1, f=1\};F=2,m_F=-2 \rangle$} & \multicolumn{1}{c}{$\pi$} &\multicolumn{1}{c}{0.89(1)}& \multicolumn{1}{c}{1.09} & \multicolumn{1}{c}{30} & \multicolumn{1}{c}{299.8698(2)} & \multicolumn{1}{c}{0.0064(8)}  \\ \hline

\multicolumn{1}{c}{$\vert \{f=1, f=1\};F=2,m_F=-1 \rangle$} & \multicolumn{1}{c}{$\sigma_+$} &\multicolumn{1}{c}{0.89(1)}& \multicolumn{1}{c}{1.09} & \multicolumn{1}{c}{30} & \multicolumn{1}{c}{300.4173(9)} & \multicolumn{1}{c}{0.0426(28)}  \\ \hline

\multicolumn{1}{c}{$\vert \{f=1, f=2\};F=2, m_F=-2,-1 \rangle$} & \multicolumn{1}{c}{$\sigma_+,\pi$} &\multicolumn{1}{c}{0.89(1)}& \multicolumn{1}{c}{5.33} & \multicolumn{1}{c}{80} & \multicolumn{1}{c}{1254.1(36)} & \multicolumn{1}{c}{172(22)}  \\ \hline

\multicolumn{1}{c}{$\vert \{f=1, f=2\};F=1,m_F=-1 \rangle$} & \multicolumn{1}{c}{$\sigma_+$} &\multicolumn{1}{c}{0.89(1)}& \multicolumn{1}{c}{0.66} & \multicolumn{1}{c}{30} & \multicolumn{1}{c}{1561.2835(31)} & \multicolumn{1}{c}{0.063(11)}  \\ \hline

\multicolumn{1}{c}{$\vert \{f=1, f=2\};F=3,m_F=-3 \rangle$} & \multicolumn{1}{c}{$\sigma_-$} &\multicolumn{1}{c}{0.89(1)}& \multicolumn{1}{c}{0.66} & \multicolumn{1}{c}{30} & \multicolumn{1}{c}{1561.9146(37)} & \multicolumn{1}{c}{0.241(15)}  \\ \hline

\multicolumn{1}{c}{$\vert \{f=1, f=2\};F=3,m_F=-2 \rangle$} & \multicolumn{1}{c}{$\pi$} &\multicolumn{1}{c}{0.89(1)}& \multicolumn{1}{c}{0.66} & \multicolumn{1}{c}{30} & \multicolumn{1}{c}{1562.1311(25)} & \multicolumn{1}{c}{0.0356(90)}  \\ \hline

\multicolumn{1}{c}{$\vert \{f=1, f=2\};F=3,m_F=-1 \rangle$} & \multicolumn{1}{c}{$\sigma_+$} &\multicolumn{1}{c}{0.89(1)}& \multicolumn{1}{c}{0.66} & \multicolumn{1}{c}{30} & \multicolumn{1}{c}{1562.3420(15)} & \multicolumn{1}{c}{0.0746(52)}  \\ \hline

\hline\hline
\end{tabular}
\caption{Experimental parameters and fit results for the microwave photoassociation spectra of Fig.~\ref{fig:spectro}(a) and Fig.~\ref{fig:spectrof_1_f_1}. For each probed molecular state, we indicate the polarization of the microwave photon involved in the transition from the initial atomic state, the amplitude of the static magnetic field $B_s$ and of the $\sigma^-$ component of the microwave field $|B_-|$, the duration of the microwave pulse $\tau$, and the fitted frequency of the microwave field at resonance $\omega_0$ and its corresponding FWHM.
\label{tab:one_photon_trans}
}
\end{table*}

\begin{table*}[htb]
\centering
\begin{tabular}{lllllll} 

\hline\hline
 \multicolumn{1}{c}{Final main spin state} & Polarization & \multicolumn{1}{c}{$B_{\textrm{s}}$~(G)} & $|B_-|$~(G) & $\tau$~(ms) & \multicolumn{1}{c}{$\omega_0/2\pi$~(MHz)} & FWHM~(MHz) \\ \hline
 
 \multicolumn{1}{c}{$\vert \{f=1, f=2\};F=1,m_F=-1 \rangle$} & \multicolumn{1}{c}{$\sigma_+$} & \multicolumn{1}{c}{0.88(1)} & \multicolumn{1}{c}{0.83} & \multicolumn{1}{c}{30} & \multicolumn{1}{c}{1561.328(2)}&
  \multicolumn{1}{c}{0.061(17)}\\
 
  \multicolumn{1}{c}{$\vert \{f=1, f=2\};F=3,m_F=-3 \rangle$} & \multicolumn{1}{c}{$\sigma_-$} & \multicolumn{1}{c}{0.88(1)} & \multicolumn{1}{c}{0.33} & \multicolumn{1}{c}{10} & \multicolumn{1}{c}{1561.942(1)}&
  \multicolumn{1}{c}{0.068(5)}\\
  
  \multicolumn{1}{c}{$\vert \{f=1, f=2\};F=3,m_F=-2 \rangle$} & \multicolumn{1}{c}{$\pi$} & \multicolumn{1}{c}{0.88(1)} & \multicolumn{1}{c}{0.83} & \multicolumn{1}{c}{30} & \multicolumn{1}{c}{1562.158(1)}&
  \multicolumn{1}{c}{0.056(20)}\\
  
   \multicolumn{1}{c}{$\vert \{f=1, f=2\};F=3,m_F=-1 \rangle$} & \multicolumn{1}{c}{$\sigma_+$} & \multicolumn{1}{c}{0.88(1)} & \multicolumn{1}{c}{0.59} & \multicolumn{1}{c}{20} & \multicolumn{1}{c}{1562.362(1)}&
 \multicolumn{1}{c}{0.045(4)}\\
  
  \hline \hline
  
    \multicolumn{1}{c}{$\vert \{f=1, f=1\};F=2,m_F=-2 \rangle$} & \multicolumn{1}{c}{$\pi$} & \multicolumn{1}{c}{1.88(1)} & \multicolumn{1}{c}{1.09} & \multicolumn{1}{c}{30} & \multicolumn{1}{c}{300.0683(2)} & \multicolumn{1}{c}{0.0029(5)}\\
  
    \multicolumn{1}{c}{$\vert \{f=1, f=1\};F=2,m_F=-1 \rangle$} & \multicolumn{1}{c}{$\sigma_+$} & \multicolumn{1}{c}{1.88(1)} & \multicolumn{1}{c}{1.09} & \multicolumn{1}{c}{30} & \multicolumn{1}{c}{301.2037(9)} &\multicolumn{1}{c}{0.031(4)} \\
 
   \multicolumn{1}{c}{$\vert \{f=1, f=2\};F=1,m_F=-1 \rangle$} & \multicolumn{1}{c}{$\sigma_+$} & \multicolumn{1}{c}{1.88(1)} & \multicolumn{1}{c}{0.83} & \multicolumn{1}{c}{100} & \multicolumn{1}{c}{1558.508(2)}&
  \multicolumn{1}{c}{0.044(6)}\\
 
  \multicolumn{1}{c}{$\vert \{f=1, f=2\};F=3,m_F=-3 \rangle$} & \multicolumn{1}{c}{$\sigma_-$} & \multicolumn{1}{c}{1.88(1)} & \multicolumn{1}{c}{0.34} & \multicolumn{1}{c}{10} & \multicolumn{1}{c}{1559.833(1)}&
  \multicolumn{1}{c}{0.050(4)}\\
  
  \multicolumn{1}{c}{$\vert \{f=1, f=2\};F=3,m_F=-2 \rangle$} & \multicolumn{1}{c}{$\pi$} & \multicolumn{1}{c}{1.88(1)} & \multicolumn{1}{c}{0.66} & \multicolumn{1}{c}{200} & \multicolumn{1}{c}{1560.281(1)}&
  \multicolumn{1}{c}{0.029(9)}\\
  
    \multicolumn{1}{c}{$\vert \{f=1, f=2\};F=3,m_F=-1 \rangle$} & \multicolumn{1}{c}{$\sigma_+$} & \multicolumn{1}{c}{1.88(1)} & \multicolumn{1}{c}{0.60} & \multicolumn{1}{c}{20} & \multicolumn{1}{c}{1560.731(1)}&
 \multicolumn{1}{c}{0.043(4)}\\
  
  \hline \hline
  
 \multicolumn{1}{c}{$\vert \{f=1, f=2\};F=1,m_F=-1 \rangle$} & \multicolumn{1}{c}{$\sigma_+$} & \multicolumn{1}{c}{2.43(1)} & \multicolumn{1}{c}{0.60} & \multicolumn{1}{c}{100} & \multicolumn{1}{c}{1556.980(2)} & \multicolumn{1}{c}{0.048(7)}\\
 
  \multicolumn{1}{c}{$\vert \{f=1, f=2\};F=3,m_F=-3 \rangle$} & \multicolumn{1}{c}{$\sigma_-$} & \multicolumn{1}{c}{2.43(1)} & \multicolumn{1}{c}{0.34} & \multicolumn{1}{c}{10} & \multicolumn{1}{c}{1558.694(1)} & \multicolumn{1}{c}{0.048(3)}\\
  
  \multicolumn{1}{c}{$\vert \{f=1, f=2\};F=3,m_F=-2 \rangle$} & \multicolumn{1}{c}{$\pi$} & \multicolumn{1}{c}{2.43(1)} & \multicolumn{1}{c}{0.34} & \multicolumn{1}{c}{200} & \multicolumn{1}{c}{1559.272(4)} & \multicolumn{1}{c}{0.045(13)}\\
  
   \multicolumn{1}{c}{$\vert \{f=1, f=2\};F=3,m_F=-1 \rangle$} & \multicolumn{1}{c}{$\sigma_+$} & \multicolumn{1}{c}{2.43(1)} & \multicolumn{1}{c}{0.34} & \multicolumn{1}{c}{20} & \multicolumn{1}{c}{1559.852(2)} &\multicolumn{1}{c}{0.035(4)} \\
  
  \hline \hline
  
    \multicolumn{1}{c}{$\vert \{f=1, f=1\};F=2,m_F=-2 \rangle$} & \multicolumn{1}{c}{$\pi$} & \multicolumn{1}{c}{2.93(1)} & \multicolumn{1}{c}{3.10} & \multicolumn{1}{c}{30} & \multicolumn{1}{c}{300.2769(8)} & \multicolumn{1}{c}{0.008(2)}\\
    
      \multicolumn{1}{c}{$\vert \{f=1, f=1\};F=2,m_F=-1 \rangle$} & \multicolumn{1}{c}{$\sigma_+$} & \multicolumn{1}{c}{2.93(1)} & \multicolumn{1}{c}{1.09} & \multicolumn{1}{c}{30} & \multicolumn{1}{c}{302.046(1)} & \multicolumn{1}{c}{0.031(5)}\\
 
 \multicolumn{1}{c}{$\vert \{f=1, f=2\};F=1,m_F=-1 \rangle$} & \multicolumn{1}{c}{$\sigma_+$} & \multicolumn{1}{c}{2.93(1)} & \multicolumn{1}{c}{0.84} & \multicolumn{1}{c}{100} & \multicolumn{1}{c}{1555.594(4)} & \multicolumn{1}{c}{0.026(16)}\\
 
  \multicolumn{1}{c}{$\vert \{f=1, f=2\};F=3,m_F=-3 \rangle$} & \multicolumn{1}{c}{$\sigma_-$} & \multicolumn{1}{c}{2.93(1)} & \multicolumn{1}{c}{0.34} & \multicolumn{1}{c}{10} & \multicolumn{1}{c}{1557.628(1)} & \multicolumn{1}{c}{0.028(4)}\\
  
  \multicolumn{1}{c}{$\vert \{f=1, f=2\};F=3,m_F=-2 \rangle$} & \multicolumn{1}{c}{$\pi$} & \multicolumn{1}{c}{2.93(1)} & \multicolumn{1}{c}{0.83} & \multicolumn{1}{c}{200} & \multicolumn{1}{c}{1558.354(1)} & \multicolumn{1}{c}{0.022(11)}\\
  
  \multicolumn{1}{c}{$\vert \{f=1, f=2\};F=3,m_F=-1 \rangle$} & \multicolumn{1}{c}{$\sigma_+$} & \multicolumn{1}{c}{2.93(1)} & \multicolumn{1}{c}{0.60} & \multicolumn{1}{c}{50} & \multicolumn{1}{c}{1559.038(2)} & \multicolumn{1}{c}{0.072(16)}\\
  
  \hline \hline
  
    \multicolumn{1}{c}{$\vert \{f=1, f=1\};F=2,m_F=-1 \rangle$} & \multicolumn{1}{c}{$\sigma_+$} & \multicolumn{1}{c}{4.57(1)} & \multicolumn{1}{c}{1.09} & \multicolumn{1}{c}{30} & \multicolumn{1}{c}{303.358(1)} & \multicolumn{1}{c}{0.023(5)}\\
 
  \multicolumn{1}{c}{$\vert \{f=1, f=2\};F=3,m_F=-3 \rangle$} & \multicolumn{1}{c}{$\sigma_-$} & \multicolumn{1}{c}{4.57(1)} & \multicolumn{1}{c}{0.77} & \multicolumn{1}{c}{10} & \multicolumn{1}{c}{1554.205(3)} &\multicolumn{1}{c}{0.020(13)} \\
  
  \multicolumn{1}{c}{$\vert \{f=1, f=2\};F=3,m_F=-1 \rangle$} & \multicolumn{1}{c}{$\sigma_+$} & \multicolumn{1}{c}{4.57(1)} & \multicolumn{1}{c}{0.84} & \multicolumn{1}{c}{30} & \multicolumn{1}{c}{1556.425(2)} & \multicolumn{1}{c}{0.071(19)}\\
  
  \hline \hline
  
\end{tabular}
\caption{Experimental parameters and fit results for the microwave photoassociation spectra of Fig.~\ref{fig:energy_vs_B}. For each probed molecular state, we indicate the polarization of the microwave photon involved in the transition from the initial atomic state, the amplitude of the static magnetic field $B_s$ and of the $\sigma^-$ component of the microwave field $|B_-|$, the duration of the microwave pulse $\tau$, and the fitted frequency of the microwave field at resonance $\omega_0$ and its corresponding FWHM.
\label{manon_data}
}
\end{table*}

\begin{table*}[htb]
\centering
\begin{tabular}{llllll} 

\hline\hline
 \multicolumn{1}{c}{Final main spin state} & Polarization & $|B_-|$~(G) & $\tau$~(ms) & \multicolumn{1}{c}{$\omega_0/2\pi$~(MHz)} & \multicolumn{1}{c}{FWHM~(MHz)} \\ \hline
 
 \multicolumn{1}{c}{$\vert \{f=1, f=1\};F=2,m_F=-2 \rangle$} & \multicolumn{1}{c}{[$\sigma_+$,$\sigma_+$], [$\sigma_-$,$\sigma_-$]} & \multicolumn{1}{c}{8.27} & \multicolumn{1}{c}{1 } & \multicolumn{1}{c}{1623.855(10)} & \multicolumn{1}{c}{0.59(4)}  \\ 
 
  \multicolumn{1}{c}{$\vert \{f=1, f=1\};F=2,m_F=-2 \rangle$} & \multicolumn{1}{c}{[$\sigma_+$,$\sigma_+$], [$\sigma_-$,$\sigma_-$]} & \multicolumn{1}{c}{7.42} & \multicolumn{1}{c}{2} & \multicolumn{1}{c}{1623.555(20)} & \multicolumn{1}{c}{0.77(10)}  \\ 
  
  \multicolumn{1}{c}{$\vert \{f=1, f=1\};F=2,m_F=-2 \rangle$} & \multicolumn{1}{c}{[$\sigma_+$,$\sigma_+$], [$\sigma_-$,$\sigma_-$]} & \multicolumn{1}{c}{6.61} & \multicolumn{1}{c}{2} & \multicolumn{1}{c}{1623.268(9)} & \multicolumn{1}{c}{0.437(35)}  \\ 
  
    \multicolumn{1}{c}{$\vert \{f=1, f=1\};F=2,m_F=-2 \rangle$} & \multicolumn{1}{c}{[$\sigma_+$,$\sigma_+$], [$\sigma_-$,$\sigma_-$]} & \multicolumn{1}{c}{4.12} & \multicolumn{1}{c}{4} & \multicolumn{1}{c}{1622.671(10)} & \multicolumn{1}{c}{0.11(4)}  \\ 
 \hline
  \multicolumn{1}{c}{$\vert \{f=2, f=2\};F=4,m_F=-4 \rangle$} & \multicolumn{1}{c}{[$\sigma_-$,$\sigma_-$]} & \multicolumn{1}{c}{8.21} & \multicolumn{1}{c}{10} & \multicolumn{1}{c}{1665.770(1)} & \multicolumn{1}{c}{0.059(4)}  \\
  
    \multicolumn{1}{c}{$\vert \{f=2, f=2\};F=4,m_F=-4 \rangle$} & \multicolumn{1}{c}{[$\sigma_-$,$\sigma_-$]} & \multicolumn{1}{c}{6.56} & \multicolumn{1}{c}{10} & \multicolumn{1}{c}{1665.786(1)} & \multicolumn{1}{c}{0.061(4) }  \\
    
    \multicolumn{1}{c}{$\vert \{f=2, f=2\};F=4,m_F=-4 \rangle$} & \multicolumn{1}{c}{[$\sigma_-$,$\sigma_-$]} & \multicolumn{1}{c}{4.92} & \multicolumn{1}{c}{10} & \multicolumn{1}{c}{1665.788(1)} & \multicolumn{1}{c}{0.047(4)}  \\ \hline
 
   \multicolumn{1}{c}{$\vert \{f=2, f=2\};F=4,m_F=-2 \rangle$} & \multicolumn{1}{c}{[$\sigma_+$,$\sigma_-$]} & \multicolumn{1}{c}{8.21} & \multicolumn{1}{c}{10} & \multicolumn{1}{c}{1666.4251(5)} & \multicolumn{1}{c}{0.029(2)} \\
   
 \multicolumn{1}{c}{$\vert \{f=2, f=2\};F=4,m_F=-2 \rangle$} & \multicolumn{1}{c}{[$\sigma_+$,$\sigma_-$]} & \multicolumn{1}{c}{7.39} & \multicolumn{1}{c}{10} & \multicolumn{1}{c}{1666.4235(7)} & \multicolumn{1}{c}{0.038(4)} \\ 
   
  \multicolumn{1}{c}{$\vert \{f=2, f=2\};F=4,m_F=-2 \rangle$} & \multicolumn{1}{c}{[$\sigma_+$,$\sigma_-$]} & \multicolumn{1}{c}{6.57} & \multicolumn{1}{c}{10} & \multicolumn{1}{c}{1666.4219(8)} & \multicolumn{1}{c}{0.028(3)} \\ \hline
 
    \multicolumn{1}{c}{$\vert \{f=2, f=2\};F=4,m_F=0 \rangle$} & \multicolumn{1}{c}{[$\sigma_+$,$\sigma_+$]} & \multicolumn{1}{c}{8.22} & \multicolumn{1}{c}{10} & \multicolumn{1}{c}{1667.0576(4)} & \multicolumn{1}{c}{0.032(2)}\\
 
\hline \hline
\end{tabular}
\caption{Experimental parameters and fit results for the microwave photoassociation spectra of Fig.~\ref{fig:energy_vs_Bmw}. For each probed molecular state, we indicate the polarization of the microwave photons involved in the transition from the initial atomic state, the amplitude of the $\sigma^-$ component of the microwave field $|B_-|$, the duration of the microwave pulse $\tau$, and the fitted frequency of the microwave field at resonance $\omega_0$ and its corresponding FWHM. For all these data, the amplitude of the static magnetic field $B_s$ is 0.92(1)\,G. Note that the energy $\Delta E_0$ of the corresponding molecular bound states can be deduced from $\omega_0$. For the $F=4$ states, it is simply $\Delta E_0/\hbar=2\omega_0$ while for the $F=2$ states, it depends on the frequency of the microwave compensation field, $\Delta E_0/\hbar=\omega_0-\omega_\textrm{c}$, with $\omega_c=2\pi\times$\SI{1922.128128}{\mega\hertz}.
\label{tab:2photons}
}
\end{table*}

\begin{table*}[htb]
\centering
\begin{tabular}{llllll} 

\hline\hline
 \multicolumn{1}{c}{Final main spin state} & Polarization & $|B_-|$~(G) & $\tau$ (ms) & \multicolumn{1}{c}{$\omega_0/2\pi$~(MHz)} & \multicolumn{1}{c}{FWHM~(MHz)} \\ \hline

\multicolumn{1}{c}{$\vert \{f=1, f=2\};F=1,m_F=-1 \rangle$} & \multicolumn{1}{c}{$\sigma_+$} & \multicolumn{1}{c}{2.50} & \multicolumn{1}{c}{10} & \multicolumn{1}{c}{1561.051(2)} & \multicolumn{1}{c}{0.222(17)}  \\ \hline

\multicolumn{1}{c}{$\vert \{f=1, f=2\};F=1,m_F=-1 \rangle$} & \multicolumn{1}{c}{$\sigma_+$} & \multicolumn{1}{c}{4.17} & \multicolumn{1}{c}{10} & \multicolumn{1}{c}{1560.775(7)} & \multicolumn{1}{c}{0.477(60)}  \\ \hline

\multicolumn{1}{c}{$\vert \{f=1, f=2\};F=1,m_F=-1 \rangle$} & \multicolumn{1}{c}{$\sigma_+$} & \multicolumn{1}{c}{5.86} & \multicolumn{1}{c}{5} & \multicolumn{1}{c}{1560.271(61)} & \multicolumn{1}{c}{1.20(88)}  \\ \hline \hline

\multicolumn{1}{c}{$\vert \{f=1, f=2\};F=3,m_F=-3 \rangle$} & \multicolumn{1}{c}{$\sigma_-$} & \multicolumn{1}{c}{2.50} & \multicolumn{1}{c}{0.25} & \multicolumn{1}{c}{1561.831(1)} & \multicolumn{1}{c}{0.104(5)}  \\ \hline

\multicolumn{1}{c}{$\vert \{f=1, f=2\};F=3,m_F=-3 \rangle$} & \multicolumn{1}{c}{$\sigma_-$} & \multicolumn{1}{c}{4.17} & \multicolumn{1}{c}{0.1} & \multicolumn{1}{c}{1561.701(1)} & \multicolumn{1}{c}{0.129(6)}  \\ \hline

\multicolumn{1}{c}{$\vert \{f=1, f=2\};F=3,m_F=-3 \rangle$} & \multicolumn{1}{c}{$\sigma_-$} & \multicolumn{1}{c}{5.85} & \multicolumn{1}{c}{0.1} & \multicolumn{1}{c}{1561.507(2)} & \multicolumn{1}{c}{0.229(11)}  \\ \hline

\multicolumn{1}{c}{$\vert \{f=1, f=2\};F=3,m_F=-3 \rangle$} & \multicolumn{1}{c}{$\sigma_-$} & \multicolumn{1}{c}{7.52} & \multicolumn{1}{c}{0.1} & \multicolumn{1}{c}{1561.172(6)} & \multicolumn{1}{c}{0.399(24)}  \\ \hline

\multicolumn{1}{c}{$\vert \{f=1, f=2\};F=3,m_F=-3 \rangle$} & \multicolumn{1}{c}{$\sigma_-$} & \multicolumn{1}{c}{8.35} & \multicolumn{1}{c}{0.1} & \multicolumn{1}{c}{1561.103(5)} & \multicolumn{1}{c}{0.521(28)}  \\ \hline\hline

\multicolumn{1}{c}{$\vert \{f=1, f=2\};F=3,m_F=-1 \rangle$} & \multicolumn{1}{c}{$\sigma_+$} & \multicolumn{1}{c}{2.50} & \multicolumn{1}{c}{10} & \multicolumn{1}{c}{1562.265(1)} & \multicolumn{1}{c}{0.111(4)}  \\ \hline

\multicolumn{1}{c}{$\vert \{f=1, f=2\};F=3,m_F=-1 \rangle$} & \multicolumn{1}{c}{$\sigma_+$} & \multicolumn{1}{c}{4.16} & \multicolumn{1}{c}{5} & \multicolumn{1}{c}{1562.197(2)} & \multicolumn{1}{c}{0.122(7)}  \\ \hline

\multicolumn{1}{c}{$\vert \{f=1, f=2\};F=3,m_F=-1 \rangle$} & \multicolumn{1}{c}{$\sigma_+$} & \multicolumn{1}{c}{5.83} & \multicolumn{1}{c}{3} & \multicolumn{1}{c}{1562.098(2)} & \multicolumn{1}{c}{0.116(10)}  \\ \hline

\multicolumn{1}{c}{$\vert \{f=1, f=2\};F=3,m_F=-1 \rangle$} & \multicolumn{1}{c}{$\sigma_+$} & \multicolumn{1}{c}{6.67} & \multicolumn{1}{c}{2} & \multicolumn{1}{c}{1562.040(4)} & \multicolumn{1}{c}{0.124(18)}  \\ \hline\hline

\end{tabular}
\caption{Experimental parameters and fit results for the microwave photoassociation spectra of Fig.~\ref{fig:f_1_f_2_F_1_F_3_vs_Bmw_2_photons}. For each probed molecular state, we indicate the polarization of the microwave photon involved in the transition from the initial atomic state, the amplitude of the $\sigma^-$ component of the microwave field $|B_-|$, the duration of the microwave pulse $\tau$, and the fitted frequency of the microwave field at resonance $\omega_0$ and its corresponding FWHM. For all these data, the amplitude of the static magnetic field $B_s$ is 0.90(1)\,G.
\label{tab:dressing}
}
\end{table*}

\clearpage


%

\end{document}